\documentclass[
 aps,
 prl,
 amsmath,amssymb,
 reprint,
 superscriptaddress,
 floatfix
]{revtex4-2}

\usepackage[english]{babel}
\usepackage[utf8]{inputenc}
\usepackage[T1]{fontenc}
\usepackage{mathrsfs}
\usepackage{bbm}
\usepackage{bm}
\usepackage{enumerate}
\usepackage{graphicx}
\usepackage[dvipsnames]{xcolor}
\usepackage{subdepth}
\usepackage{authoraftertitle}
\usepackage{physics}
\usepackage{lipsum}
\usepackage{comment}

\usepackage[colorlinks,
linkcolor=BrickRed,
citecolor=MidnightBlue,
urlcolor=MidnightBlue,
bookmarks=true,        
bookmarksopen=true,    
bookmarksnumbered=true,
]{hyperref}

\usepackage[raggedright,small]{titlesec}

\renewcommand*\d{\mathop{}\!\mathrm{d}}
\newcommand{\dirac}[1]{\,\delta\!\left(#1\right)}
\newcommand{\heav}[1]{\,\Theta\!\left(#1\right)}

\newcommand{\av}[1]{\left\langle#1\right\rangle}

\newcommand{\conj}[1]{\bar{#1}}
\newcommand{\pd}{\partial}

\renewcommand{\i}{\mathrm{i}}

\newcommand{\sL}{\mathcal{L}}

\newcommand{\sD}{\mathcal{D}}
\newcommand{\sT}{\mathcal{T}}
\newcommand{\sC}{\mathcal{C}}
\newcommand{\sP}{\mathcal{P}}

\renewcommand{\(}{\left(}
\renewcommand{\)}{\right)}
\newcommand{\id}{\mathbbm{1}}

\newcommand{\eff}{\mathrm{eff}}
\newcommand{\rmT}{\mathrm{T}}
\newcommand{\rmTb}{{\bar{\mathrm{T}}}}
\newcommand{\rmK}{\mathrm{K}}
\newcommand{\rmR}{\mathrm{R}}
\newcommand{\rmA}{\mathrm{A}}
\newcommand{\rmH}{\mathrm{H}}

\newcommand{\Omegatilde}{\widetilde{\Omega}}

\global\long\def\k{\mathbf{k}}

\global\long\def\t#1{\text{#1}}

\newcommand{\sect}[1]{\textit{#1.}---}

\newcounter{ls}

\newcounter{pr}

\newcommand{\figpath}{Figs}
\newcommand{\bibpath}{BibFiles}

\begin{document}
	
\title{Lindbladian dissipation of strongly-correlated quantum matter}
\author{Lucas  S\'a}
\email{lucas.seara.sa@tecnico.ulisboa.pt}

\affiliation{CeFEMA, Instituto Superior T\'ecnico, Universidade de Lisboa, Av.\ Rovisco Pais, 1049-001 Lisboa, Portugal}

\author{Pedro Ribeiro}
\email{ribeiro.pedro@tecnico.ulisboa.pt}

\affiliation{CeFEMA, Instituto Superior T\'ecnico, Universidade de Lisboa, Av.\ Rovisco Pais, 1049-001 Lisboa, Portugal}
\affiliation{Beijing Computational Science Research Center, Beijing 100193, China}

\author{Toma\v z Prosen}
\email{tomaz.prosen@fmf.uni-lj.si}

\affiliation{Department of Physics, Faculty of Mathematics and Physics, University of Ljubljana, Ljubljana, Slovenia}

\begin{abstract}
We propose the Sachdev-Ye-Kitaev Lindbladian as a paradigmatic solvable model of dissipative many-body quantum chaos. It describes $N$ strongly coupled Majorana fermions with random all-to-all interactions, with unitary evolution given by a quartic Hamiltonian and the coupling to the environment described by $M$ quadratic jump operators, rendering the full Lindbladian quartic in the Majorana operators. 
Analytical progress is possible by developing a dynamical mean-field theory for the Liouvillian time evolution on the Keldysh contour. By disorder-averaging the interactions, we derive an (exact) effective action for two collective fields (Green's function and self-energy). In the large-$N$, large-$M$ limit, we obtain the saddle-point equations satisfied by the collective fields, which determine the typical timescales of the dissipative evolution, particularly, the spectral gap that rules the relaxation of the system to its steady state. We solve the saddle-point equations numerically and find that, for strong or intermediate dissipation, the system relaxes exponentially, with a spectral gap that can be computed analytically, while for weak dissipation, there are oscillatory corrections to the exponential relaxation.
In this letter, we illustrate the feasibility of analytical calculations in strongly correlated dissipative quantum matter.
\end{abstract}

\maketitle

The dynamics of complex interacting many-body open quantum systems, their timescales, and their correlations are a timely topic with major conceptual and experimental significance.
In addition to dissipation and decoherence, contact with different environments may induce currents of otherwise conserved quantities, such as energy and charge, and the observables of the system typically attain a steady state. A compact form for describing the dynamics of a quantum system in the presence of an environment with a short memory time (i.e., in the Markovian approximation) is to consider the quantum master equation for the density matrix of the system, $\pd_t\rho=\sL\(\rho\)$, where the Liouvillian generator is of the Lindblad form~\cite{belavin1969,lindblad1976,gorini1976}:
\begin{equation}\label{eq:Lindblad_eq}
\sL \(\rho\)= -\i \comm{H}{\rho}+\sum_{m=1}^{M}\(2L_m\rho L_m^\dagger-\acomm{L_m^\dagger L_m}{\rho}\).
\end{equation}
Here, $H$ is the Hamiltonian of the system, while the $M$ jump operators $L_m$ describe the effective coupling to the environment.
 
Even within this simplified setup, describing interacting open quantum systems is a daunting task. Considerable progress has been achieved in the past decade for integrable models. In generic (i.e., chaotic) closed many-body quantum systems, interactions may entail such a complex structure that the Hamiltonian behaves in several aspects like a large random matrix, as conjectured by Bohigas, Giannoni, and Schmit~\cite{bohigas1984}.
Extending this result to the dissipative realm is a fundamental problem that has attracted considerable attention recently~\cite{grobe1988,grobe1989,akemann2019,sa2019csr,hamazaki2019,li2021PRL,garcia-garcia2022PRX}.
Along similar lines, the past couple of years have seen the development of the (non-Hermitian) random matrix theory of Lindbladian dynamics~\cite{denisov2018,can2019,can2019a,sa2019,wang2020,sommer2021,lange2021,tarnowski2021,li2022PRB}. By randomly sampling the Hamiltonian and jump operators, many statistical properties, including the spectral support~\cite{denisov2018} and distribution~\cite{tarnowski2021}, the spectral gap~\cite{can2019,can2019a,sa2019}, and the steady state~\cite{sa2019,sa2020} have been computed. However, physical systems have few-body interactions, rendering them very different from dense random matrices. It is natural to ask what properties are similar (i.e., universal) in both cases. Steps in this direction were taken in Refs.~\cite{wang2020,sommer2021,li2022PRB}, where local operators were modeled as Pauli strings with a fixed number of non-identity operators.

We instead propose using the Sachdev-Ye-Kitaev (SYK) model~\cite{sachdev1993,kitaev2015TALK1,kitaev2015TALK2,kitaev2015TALK3,sachdev2015PRX}, a model of $N$ Majorana fermions with random all-to-all couplings, to describe both the Hamiltonian and the jump operators of the dissipative system. The SYK model originated in nuclear physics 50 years ago~\cite{french1970PhysLettB,french1971PhysLettB,bohigas1971PhysLettB,bohigas1971PhysLettB2,mon1975AnnPhys} but has seen a recent surge of interest because of its connection to two-dimensional quantum gravity, after it was shown to be maximally chaotic, exactly-solvable at strong coupling, and near conformal~\cite{kitaev2015TALK1,kitaev2015TALK2,kitaev2015TALK3,polchinski2016JHEP,maldacena2016JHEP,maldacena2016PRD,jevicki2016JHEP,jevicki2016JHEP2,gross2017JHEP}.
Later, it was also found that it displays an exponential growth of low-energy excitations typical of black holes and heavy nuclei~\cite{garcia-garcia2017PRD,cotler2017JHEP,stanford2017JHEP}, that it realizes the full Altland-Zirnbauer classification~\cite{garcia-garcia2016PRD,garcia-garcia2017PRD,cotler2017JHEP,you2017PRB,kanazawa2017JHEP,li2017JHEP,altland2018NuclPhysB,behrends2019PRB,sun2020PRL,sa2022PRD}, and that it captures many features of non-Fermi liquids~\cite{parcollet1999PRB,song2017PRL,zhang2017PRB,gnezdilov2018PRB,can2019PRB} and wormholes~\cite{maldacena2018ARXIV,garcia-garcia2019PRD,kim2019PRX,qi2020JHEP,plugge2020PRL,lantagne-hurtubise2020PRR,sahoo2020PRR,klebanov2020JHEP,lensky2021JHEP,garcia-garcia2021PRD2,haenel2021PRB}.
These developments placed the SYK model in a prominent position at the intersection of high-energy physics, condensed matter, and quantum chaos, as one of the few analytically tractable models of both holography and strongly interacting quantum matter. 
Moreover, several experimental implementations have been proposed~\cite{danshita2017PTEP,chew2017prb,pikulin2017PRX,chen2018PRL,franz2018NatRM,wei2021PRA}, and its practical and technological relevance has been highlighted~\cite{garcia-alvarez2017PRL,luo2019npjQI,babbush2019PRA,rossini2020PRL,rosa2020JHEP,behrends2022PRL}. Finally, non-Hermitian SYK models have also started gaining traction, with studies focusing on thermodynamics and wormhole physics~\cite{garcia-garcia2021PRD,garcia-garcia2022PRL}, symmetries and universality~\cite{garcia-garcia2022PRX}, entanglement dynamics~\cite{liu2021SciPost,zhang2021Quantum}, and the effect of decoherence on quantum chaos~\cite{xu2020,cornelius2022}.

In this letter, we exploit the solvability of the SYK model and develop an analytic theory for the relaxation of generic strongly interacting dissipative quantum systems.
Nonequilibrium real-time Hamiltonian dynamics of the SYK model (e.g., thermalization and transport) have been studied before by either coupling it to an external bath~\cite{chen2017JHEP,zhang2019PRB,almheiri2019ARXIV,zhou2020PRB,maldacena2021JHEP,cheipesh2021PRB,haldar2020PRR,zanoci2022PRR} or quenching its interactions~\cite{eberlein2017PRB,bhattacharya2019JHEP,kuhlenkamp2020PRL,larzul2022PRB,louw2022PRB} (see also Refs.~\cite{su2021JHEP,chen2021JHEP} for non-Markovian entropy dynamics and Refs.~\cite{jian2021PRL,jian2021ARXIV,altland2021ARXIV} for a continuously monitored SYK model), but a fully fledged quantum-master-equation approach to strongly interacting dissipative dynamics has remained unaddressed. Here, we bridge this gap.
Working on the Keldsyh contour~\cite{keldysh1965,kamenev2011,stefanucci2013,grozdanov2015PRD}, we extend the dynamical mean-field theory for the collective degrees of freedom (mean-field Green's function and self-energy)~\cite{sachdev2015PRX,bagrets2016NuclPhysB,bagrets2017NuclPhysB,kitaev2018JHEP} to the Lindbladian evolution. Because the interactions are random and all to all, this mean-field theory is exact. From its saddle-point equations, we compute the retarded Green's function and determine the approach to the nonequilibrium steady state.

To start, we consider the Hamiltonian $H$ and the $M$ jump operators $L_m$ in Eq.~(\ref{eq:Lindblad_eq}) to be SYK operators:
\begin{equation}\label{eq:H_L_SYK_def}
H=\sum_{i<j<k<l}^{N}J_{ijkl}\chi_i\chi_j\chi_k\chi_l
\quad \text{and} \quad
L_m=\i\sum_{i<j}^{N}\ell_{m,ij}\chi_i\chi_j.
\end{equation}
The Majorana operators $\chi_i$ satisfy the $N$-dimensional Clifford algebra $\acomm{\chi_i}{\chi_j}=\delta_{ij}$,
and the totally antisymmetric couplings $J_{ijkl}$ and $\ell_{m,ij}$ are independent Gaussian random variables with zero mean and variance
\begin{equation}
    \label{eq:J_l_moments}
    \av{J_{ijkl}^2}=\frac{3!J^2}{N^3}
    \quad\text{and}\quad
    \av{\abs{\ell_{m,ij}}^2}=\frac{\gamma^2}{N^2},
\end{equation}
respectively. ($J_{ijkl}$ must be real to ensure Hermiticity of the Hamiltonian, while $\ell_{m,ij}$ can generally be complex.)
Notice the nontrivial scaling of the quadratic SYK couplings, which is required for a nontrivial theory in the large-$N$ limit.
The scales $J$ and $\gamma$ measure the strength of the unitary and dissipative contributions to the Liouvillian, respectively.
The Hamiltonian describes coherent long-range four-body interactions, while each $L_m$ gives an independent channel for incoherent two-body interactions, in such a way that the full Liouvillian is quartic in the Majorana operators:
\begin{equation}\label{eq:Liouv_op}
\begin{split}
&\sL(\rho)=
-\i\sum_{i<j<k<l}^N J_{ijkl}\(
\chi_i\chi_j\chi_k\chi_l\rho-\rho\chi_i\chi_j\chi_k\chi_l
\)\\
&-\sum_{\substack{i<j\\k<l}}^N
\Gamma_{ijkl}\(
2\chi_i\chi_j\rho\chi_k\chi_l
-\chi_k\chi_l\chi_i\chi_j\rho
-\rho\chi_k\chi_l\chi_i\chi_j
\),
\end{split}
\end{equation}
where we defined the positive-definite matrix:
\begin{equation}\label{eq:Gamma_def}
\Gamma_{ijkl}=\sum_{m=1}^M\ell_{m,ij}\ell_{m,kl}^*,
\end{equation}
which satisfies $\Gamma_{ijkl}=-\Gamma_{jikl}=\Gamma_{jilk}=-\Gamma_{ijlk}=\Gamma_{klij}^*$. If we let $N,M\to\infty$ with $m=M/N$ fixed, $\Gamma_{ijkl}$ also becomes Gaussian distributed. Then Eq.~(\ref{eq:J_l_moments}) implies that the mean and the variance of $\Gamma_{ijkl}$ are, respectively,
\begin{equation}\label{eq:Gamma_moments}
    \av{\Gamma_{ijij}}=\frac{m\gamma^2}{N}
    \quad\text{and}\quad
    \av{\abs{\Gamma_{ijkl}}^2}_\mathrm{con}= \frac{m\gamma^4}{N^3}.
\end{equation}

The system undergoes nonunitary time evolution toward a steady state $\rho_\infty$, satisfying $\sL(\rho_\infty)=0$. For simplicity, we restrict ourselves to Hermitian jump operators (i.e., real $\ell_{m,ij}$). In that case, the steady state is the infinite-temperature state $\rho_\infty=\id/2^{N/2}$. We are interested in the relaxation to $\rho_\infty$. To that end, we consider the retarded Green's function:
\begin{equation}\label{eq:def_GR}
    \i G^\rmR(t)\,\delta_{ij}
    =\heav{t}\av{\Tr\left[\rho_{\infty}
    \acomm{\chi_{i}(t)}{\chi_j}\right]},
\end{equation}
where $\av{\cdots}$ denotes the average over both $J_{ijkl}$ and $\Gamma_{ijkl}$ and the (Heisenberg-picture) Majorana operator $\chi_i(t)$ satisfies the adjoint Lindblad equation, $\pd_t\chi_i=\sL^\dagger(\chi_i)$. The relaxation dynamics are characterized by the late-time decay of $\i G^\rmR(t)$. An exponential decay signals a well-defined spectral gap (relaxation rate).

We now switch to the Keldysh path-integral representation of the Majorana Liouvillian (see the Supplemental Material~(SM)~\cite{SM} for a derivation and Ref.~\cite{sieberer2016} for the bosonic version).
We introduce real Grassmann fields $a_{i}(z)$ living on the closed-time contour $z\in\sC=\sC^+\cup\sC^-$, where real time runs from $-\infty$ to $+\infty$ (branch $\sC^+$) and then back again to $-\infty$ (branch $\sC^-$).
The Grassmann field $a_i(t^+)$ [$a_i(t^-)$], with $t^+\in\sC^+$ ($t^-\in\sC^-$), propagates forward (backward) in time and is the path-integral representation of a Majorana operator acting on the density matrix from the left (right). Using Eq.~(\ref{eq:Liouv_op}), we can immediately write down the partition function:
\begin{equation}\label{eq:path_integral}
Z=\int \prod_{i=1}^N\sD a_i\ e^{\i S[a_i]},
\end{equation}
where we omitted an initial-state contribution that is irrelevant for the long-time dissipative dynamics and the Lindblad-Keldysh action is
\begin{equation}\label{eq:C_Keldysh_action}
\begin{split}
\i S[a_i]&=
\i\int_\sC \d z\,\frac{1}{2} \sum_{i=1}^N
a_{i}(z)\,\i\pd_za_{i}(z)
\\
&-\i \int_\sC \d z\sum_{i<j<k<l}^N
J_{ijkl}a_i(z)a_j(z)a_k(z)a_l(z)
\\
&+\int_\sC \d z \d z'\,K(z,z')\sum_{\substack{i<j\\k<l}}^N
\Gamma_{ijkl} a_i(z)a_j(z)a_k(z')a_l(z').
\end{split}
\end{equation}
The memory kernel $K(z,z')$ allows for both Markovian and non-Markovian dissipative dynamics (see the SM~\cite{SM} for a discussion on how a non-Markovian thermal bath fits into our framework). Comparing Eqs.~(\ref{eq:Liouv_op})~and~(\ref{eq:C_Keldysh_action}) we can read off the lesser and greater components of the Markovian kernel:
\begin{align}
\label{eq:K_kernel_<}
K^<(t_1,t_2)&=K(t_1^+,t_2^-)=2\dirac{t_1-t_2},
\\
\label{eq:K_kernel_>}
K^>(t_1,t_2)&=K(t_1^-,t_2^+)=0,
\end{align}
respectively.
Equations~(\ref{eq:C_Keldysh_action})--(\ref{eq:K_kernel_>}) can also be derived microscopically by tracing out the environment in a unitary system-plus-environment theory~\cite{SM}.
We further define the (mean-field) Green's function,
\begin{equation}\label{eq:G_def}
G(z,z')=-\frac{\i}{N}\sum_{i=1}^N  a_{i}(z) a_{i}(z'),
\end{equation}
and the self-energy $\Sigma(z,z')$ as a Lagrange multiplier enforcing the definition of Eq.~(\ref{eq:G_def}) in the path integral.

To proceed, we average over the random couplings $J_{ijkl}$ and $\Gamma_{ijkl}$, in the limit $N,M\to\infty$ with $m=M/N$ fixed. The averaging procedure is straightforward and is presented in the SM~\cite{SM}. The resulting averaged partition function is
\begin{equation}
\av{Z}=\int \sD G\, \sD\Sigma \ e^{\i S_\eff[G,\Sigma]},
\end{equation}
with mean-field action:
\begin{widetext}
\begin{equation}\label{eq:C_S_eff}
\begin{split}
\i &S_\eff[G,\Sigma]=
\,\frac{N}{2}\Bigg\{
\Tr\log\(\i\pd-\Sigma\)
-\int_\sC \d z \d z'\, \Sigma(z,z')\,G(z,z')
-\frac{J^2}{4}\int_\sC \d z\d z'\, \left[G(z,z')\right]^4
\\
&+\frac{m\gamma^4}{4}\int_\sC \d z \d z' \d w \d w' \,
K(z,z')\,K(w,w')
\left[G(z,w')\right]^2 \left[G(z',w)\right]^2
+m\gamma^2\int_\sC \d z \d z' \,
K(z,z')\left[G(z,z')\right]^2
\Bigg\}.
\end{split}
\end{equation}
Variation of Eq.~(\ref{eq:C_S_eff}) with respect to $\Sigma(z,z')$ and $G(z,z')$ (recall that both are antisymmetric in their contour indices) leads to the Schwinger-Dyson equations on $\sC$:
\begin{align}
	\label{eq:C_SD_Sigma}
	\(\i\pd-\Sigma\)\cdot &\,G=\id_{\sC},
	\\
	\begin{split}\label{eq:C_SD_G}
	\Sigma(z,z')=&-J^2\left[G(z,z')\right]^3
	+\frac{m \gamma^4}{2} G(z,z') \int_{\sC} \d w \d w' \,
	\left[K(z,w)K(w',z')+K(w,z)K(z',w')
	\right]\left[G(w,w')\right]^2
	\\
	&+m\gamma^2\left[K(z,z')+K(z',z)\right]G(z,z'),
	\end{split}
\end{align}
\end{widetext}
where Eq.~(\ref{eq:C_SD_Sigma}) is to be understood as a matrix equation, while Eq.~(\ref{eq:C_SD_G}) acts on each matrix element individually. These equations are exact for the SYK Lindbladian in the large-$N$, large-$M$ limit.

We now move back from contour times $(z,z')$ to real times $(t_1,t_2)$. For Majorana fermions, there is a single independent Green's function~\cite{eberlein2017PRB,babadi2015PRX},
say, the greater component, $G^>(t_1,t_2)=G(t_1^-,t_2^+)$, while the lesser component, $G^<(t_1,t_2)=G(t_1^+,t_2^-)$, satisfies $G^<(t_1,t_2)=-G^>(t_2,t_1)$.
Restricting Eq.~(\ref{eq:C_SD_Sigma}) to $(z,z')=(t_1^-,t_2^+)$ and using Eqs.~(\ref{eq:K_kernel_<}) and (\ref{eq:K_kernel_>}), Eq.~(\ref{eq:C_SD_G}) reads as~\cite{SM}:
\begin{equation}\label{eq:Sigma>}
\begin{split}
\Sigma^>&(t_1,t_2)=
G^>(t_1,t_2)\left\{
-J^2\left[G^>(t_1,t_2)\right]^2
\right.
\\
&\left.
-m\gamma^4\left[G^<(t_1,t_2)\right]^2
+2m\gamma^2\dirac{t_1-t_2}
\right\}.
\end{split}
\end{equation}

Next, we change variables to $t=t_1-t_2$ and $\sT=(t_1+t_2)/2$. For long times, $\sT\to\infty$, the system loses any information about its initial state and relaxes to the steady state. The Green's function $G^>$ depends now only on $t$, and we move to Fourier space with continuous frequencies $\omega$. We further perform a Keldysh rotation by defining the real quantities~\cite{ribeiro2015PRL}:
\begin{align}
\rho^{\pm}(\omega)&=-\frac{1}{2\pi \i}\(
G^>(\omega)\mp G^>(-\omega)
\),
\\\label{eq:def_rhoH}
\rho^\rmH(\omega)&=-\frac{1}{\pi}\sP\!\int \d \nu\, \frac{\rho^-(\nu)}{\omega-\nu},
\end{align}
and analogously for $\Sigma^>(\omega)$. Here, $\rho^+(\omega)$ is proportional to the Keldysh component of the Green's function, $G^\rmK(\omega)=-2\pi\i \rho^+(\omega)$, while the spectral function $\rho^-(\omega)$ is normalized, $\int \d \omega\, \rho^{-}(\omega)=1$, and, together with its Hilbert transform ${\rho^\rmH}(\omega)$, completely determines the retarded Green's function, $G^\rmR(\omega)=-\pi\left[\rho^\rmH(\omega)+\i\rho^-(\omega)\right]$. Because the steady state is the infinite-temperature state, we have $\rho^+(\omega)=0$, and we can write Eq.~(\ref{eq:def_GR}) as
\begin{equation}\label{eq:GR_rho-}
    \i G^\rmR(t)=\heav{t}\int \d \omega\,  \rho^-(\omega)\,\cos{\omega t}.
\end{equation}
After the Fourier transformation and Keldysh rotation, the self-energy, Eq.~(\ref{eq:Sigma>}), is given by
\begin{equation}\label{eq:SDE_sigma+-}
\begin{split}
\sigma^-&(\omega)
=\frac{m\gamma^2}{\pi}\\
+\,&\frac{J^2+m\gamma^4}{4}
\int \d\mu \d\nu\, \rho^-(\omega-\mu-\nu)\rho^-(\mu)\rho^-(\nu),
\end{split}
\end{equation}
while the Dyson equation, Eq.~(\ref{eq:C_SD_Sigma}), reads as
\begin{equation}\label{eq:SDE_rho+-}
\rho^-(\omega)=\frac{\sigma^-(\omega)}{
\left[\omega+\pi \sigma^\rmH(\omega)\right]^2
+\left[\pi \sigma^-(\omega)\right]^2}.
\end{equation}

\begin{figure}[t]
    \centering
    \includegraphics[width=0.8\columnwidth]{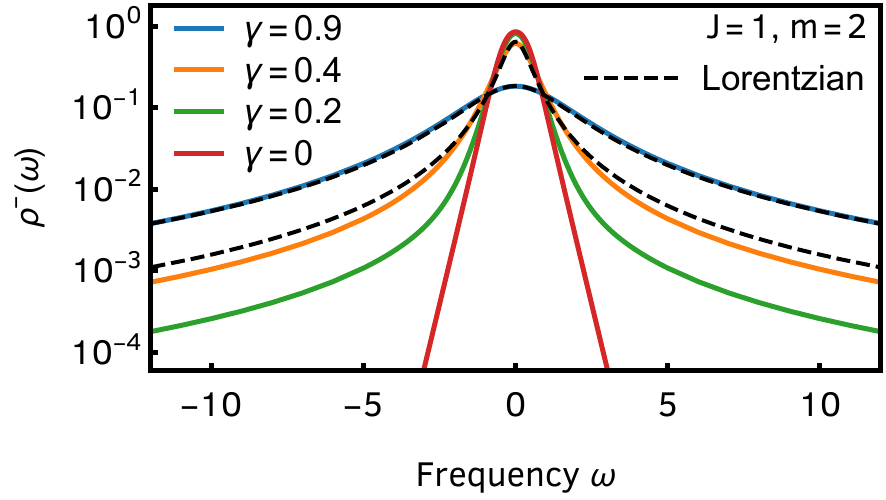}
    \caption{Spectral function~$\rho^-(\omega)$ obtained from the numerical solution of the Schwinger-Dyson equations, Eqs.~(\ref{eq:SDE_sigma+-}) and (\ref{eq:SDE_rho+-}), for $J=1$, $m=2$, and different $\gamma$. For large $\gamma$, the solution is well described by a Lorentzian (dashed lines) with the width computed analytically, Eq.~(\ref{eq:analytical_gap}). For intermediate $\gamma$ (e.g., $\gamma=0.4$), the Lorentzian ansatz still gives a reasonable description of the result (especially for low frequencies), but it fails for low dissipation.}
    \label{fig:rhominus}
\end{figure}

\begin{figure}[t]
    \centering
    \includegraphics[width=0.95\columnwidth]{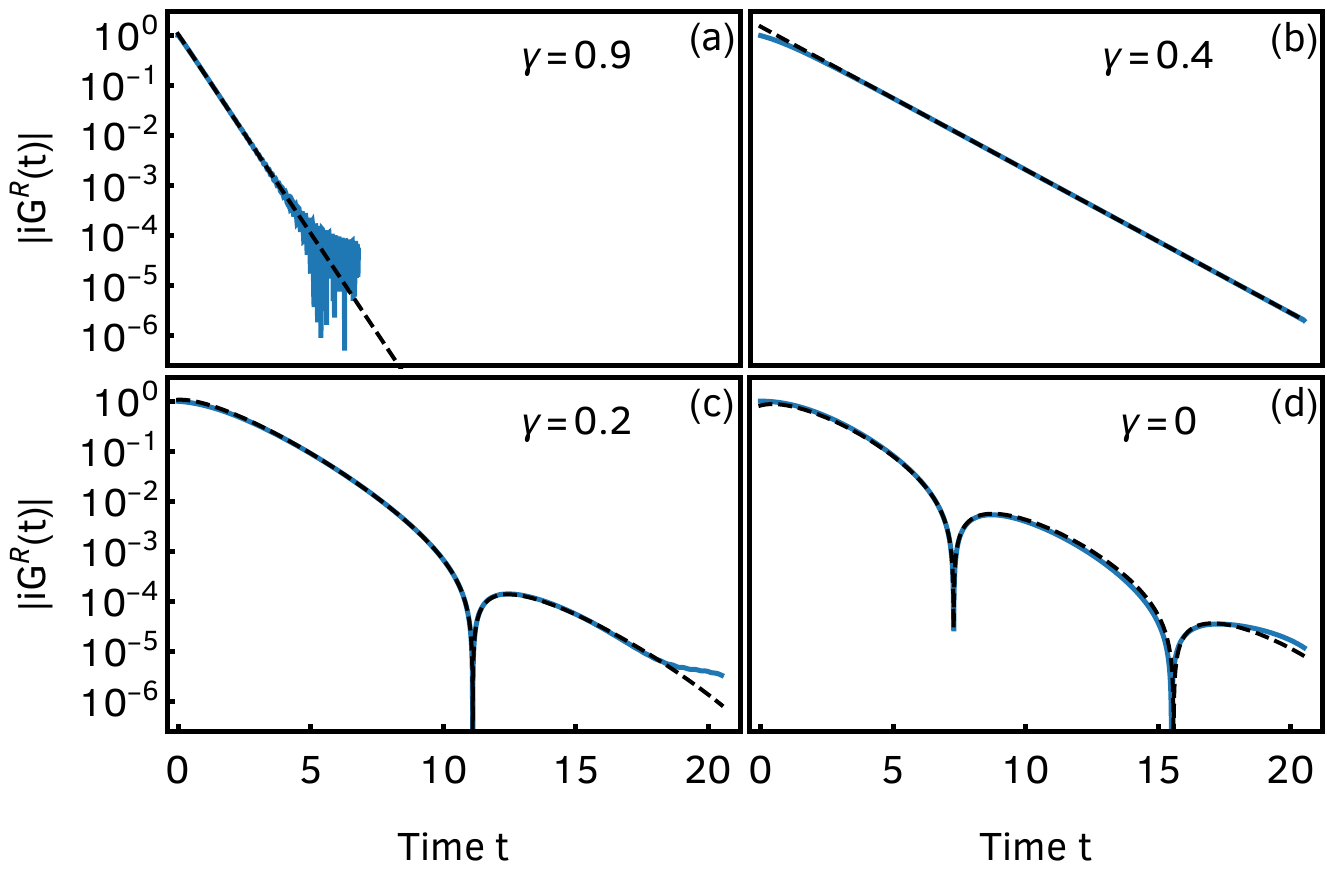}
    \caption{Retarded Green's function as a function of time for $J=1$, $m=2$, and different $\gamma$. The full blue curves are obtained by Fourier transforming the numerical solution for $\rho^-(\omega)$, as prescribed in Eq.~(\ref{eq:GR_rho-}). The dashed black lines give the best asymptotic fit to Eq.~(\ref{eq:GR_fit}).}
    \label{fig:Gret}
\end{figure}

The Schwinger-Dyson equations can be solved numerically in a self-consistent manner by proposing an ansatz for $\rho^-(\omega)$ and $\sigma^-(\omega)$ and then iterating Eqs.~(\ref{eq:SDE_sigma+-}) and (\ref{eq:SDE_rho+-}) until convergence is achieved~\cite{ribeiro2015PRL}. Details on our numerical procedure are given in the SM~\cite{SM}. The results for $J=1$, $m=2$, and different values of $\gamma$ are plotted in Fig.~\ref{fig:rhominus}. For large-enough $\gamma$, the spectral function is well approximated by a Lorentzian. Fourier transforming back to the time domain [Eq.~(\ref{eq:GR_rho-})], see Figs.~\ref{fig:Gret}(a) and \ref{fig:Gret}(b), this implies a well-defined spectral gap $\Delta$ (i.e., relaxation rate), as the retarded Green's function decays exponentially, $\i G^\rmR(t)=\heav{t}\exp\{-\Delta t\}$. The spectral gap can be determined analytically as follows. We propose the Lorentzian ansatz:
\begin{equation}\label{eq:rho-_ansatz_largegamma}
\rho^-(\omega)=\frac{1}{\pi}\frac{\Delta}{\omega^2+\Delta^2},
\end{equation}
for the spectral function, and because the Lorentzian is stable under convolution, Eq.~(\ref{eq:SDE_sigma+-}) leads to the self-energy:
\begin{equation}\label{eq:sigma-_ansatz_largegamma}
\sigma^-(\omega)=\frac{m\gamma^2}{\pi}+\frac{J^2+m\gamma^4}{4\pi}\frac{3\Delta}{\omega^2+(3\Delta)^2}.
\end{equation}
Since we are interested in the low-frequency response, we set $\omega=0$. The regime of validity of this approximation can be determined self-consistently and is presented in the SM~\cite{SM}. Plugging Eqs.~(\ref{eq:rho-_ansatz_largegamma}) and (\ref{eq:sigma-_ansatz_largegamma}) back into the Dyson equation, Eq.~(\ref{eq:SDE_rho+-}), we find
\begin{equation}\label{eq:analytical_gap}
\Delta=\frac{m\gamma^2}{2}\(
1+\sqrt{\frac{3m+1}{3m}+\frac{J^2}{3m^2\gamma^4}}
\).
\end{equation}

\begin{figure}[t]
    \centering
    \includegraphics[width=0.8\columnwidth]{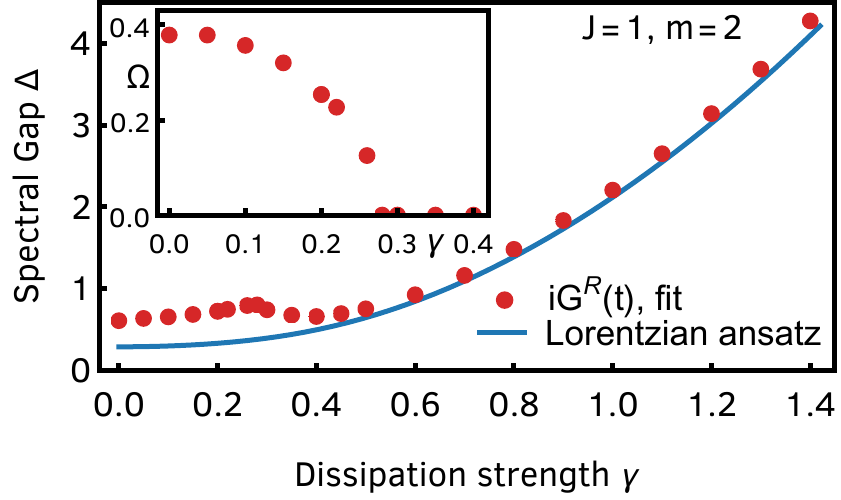}
    \caption{Spectral gap~$\Delta$ as a function of dissipation strength $\gamma$ for $J=1$ and $m=2$. The red dots are obtained from the fit to Eq.~(\ref{eq:GR_fit}), while the blue line is the analytical result from the Lorentzian ansatz, Eq.~(\ref{eq:analytical_gap}). The two agree for large $\gamma$ but saturate to different values as $\gamma\to0$. Inset: Frequency~$\Omega$ of the oscillatory correction as a function of $\gamma$. For $\gamma\gtrsim0.28$, the period of oscillations either diverges ($\Omega=0$) or becomes longer than the numerically-accessible time window.}
    \label{fig:gapPlot}
\end{figure}

Equation~(\ref{eq:analytical_gap}), the analytical relaxation rate of a strongly correlated dissipative quantum system, is the main result of this letter. The comparison with the numerical solution is given in Fig.~\ref{fig:rhominus}. For $J=1$ and $m=2$, already for $\gamma=0.9$, there is excellent agreement. Accordingly, we see a clear exponential decay of $\i G^\rmR(t)$ in Fig.~\ref{fig:Gret}~(a). For intermediate $\gamma$, say, $\gamma=0.4$, there are noticeable deviations in the tails, but the low-frequency part of $\rho^-$ is still perfectly described by Eqs.~(\ref{eq:rho-_ansatz_largegamma}) and (\ref{eq:analytical_gap}) and $\i G^\rmR(t)$ still decays exponentially, see Fig.~\ref{fig:Gret}~(b). For small $\gamma$, the tails of the spectral function are very far from Lorentzian.
This signals possible power-law or oscillatory corrections to the asymptotic decay of $G^\rmR(t)$ (depending on the precise form of $\rho^-(\omega)$, which cannot be determined analytically) that we confirm numerically, see Figs.~\ref{fig:Gret}(c) and \ref{fig:Gret}(d).
We can extract the spectral gap from $\i G^\rmR(t)$ by fitting the numerical results to an exponential function with power-law and oscillatory corrections. We found the former to be negligible in general, but the latter to be relevant for small $\gamma$, i.e., 
\begin{equation}\label{eq:GR_fit}
    \i G^\rmR(t)=A\, e^{-\Delta t}\cos(\Omega t+\phi)
\end{equation}
gives an excellent fit for $t\gg 1$
with fitting parameters $A$, $\Delta$, $\Omega$, and $\phi$. The resulting spectral gap is plotted in Fig.~\ref{fig:gapPlot} as a function of $\gamma$. We conclude that, for large $\gamma$, $\Delta$ grows quadratically, in agreement with Eq.~(\ref{eq:analytical_gap}), while it starts to deviate from the Lorentzian ansatz at intermediate values $\gamma\approx0.5$. As $\gamma$ further decreases, our results are consistent (within the numerically accessible time window) with a bifurcation of the real gap $\Delta$ to a pair of complex-conjugated gaps $\Delta\pm\i\Omega$ at $\gamma\approx0.28$, see inset of Fig.~\ref{fig:gapPlot}. Remarkably, as $\gamma\to0$, $\Delta$ saturates to a finite value, indicating that even an infinitesimally small amount of dissipation leads to relaxation at a finite rate. This is admissible given that we took the thermodynamic limit first. Notice that the strict limit $\gamma=0$ is singular, as no steady state exists, and the solution of the Schwinger-Dyson equations depends on the initial state (here, the infinite-temperature equilibrium state)~\cite{SM}. Although the Lorentzian ansatz and the numerical solution saturate to different values when $\gamma\to0$, the former still gives a qualitatively correct picture for the relaxation rate of the SYK model across all dissipation scales.

In summary, we studied the dissipative dynamics of the SYK model in the framework of the Lindbladian quantum master equation. 
We found exponential relaxation to the infinite-temperature steady state (with possible oscillatory corrections) and  analytically computed the spectral gap in the limit of strong dissipation. Our work paves the way for further analytical investigations of dissipative strongly correlated quantum matter, as many interesting questions remain unanswered. 
First, our method can be straightforwardly generalized for arbitrary $q$-body interactions~\cite{SM} (here, $q=4$). The question arises whether the physics is qualitatively the same for all $q$, and particularly, what happens in the large-$q$ limit, where there are simplifications in the standard SYK model~\cite{maldacena2016PRD}?
Second, our work can be used to study more general setups with non-Markovian dissipation by tuning the kernel $K(z,z')$. 
Third, going away from the scalings of Eq.~(\ref{eq:J_l_moments}) and considering $1/N$ corrections and non-Hermitian jump operators allows for nontrivial steady states. An analysis of the spectral and steady-state properties of general SYK Lindbladians, based on exact diagonalization along the lines of Refs.~\cite{sa2019,sa2020}, is a natural next step. 
Finally, we mention the possibility of studying the symmetries of fermionic open quantum matter~\cite{lieu2020,altland2021,garcia-garcia2022PRX} in the context of the SYK model, for which a rich classification exists in the closed case~\cite{you2017PRB,garcia-garcia2016PRD,kanazawa2017JHEP}.

\let\oldaddcontentsline\addcontentsline% Store \addcontentsline
\renewcommand{\addcontentsline}[3]{}% Make \addcontentsline a no-op

\begin{acknowledgments}
\sect{Note added}%
After our manuscript appeared on the arXiv, we were made aware of related work posted briefly after ours~\cite{kulkarni2021ARXIV}, which independently addresses the same problem through slightly different techniques. Their findings corroborate the results of Fig.~\ref{fig:gapPlot}. 

\sect{Acknowledgments}%
We are very thankful to Antonio Garc\'ia-Garc\'ia for several discussions and collaboration in related work that pointed us to the questions addressed in this letter and to Sa\v{s}o Grozdanov for comments on the manuscript. This work was supported by FCT through grants No.\ SFRH/BD/147477/2019 (LS) and UID/CTM/04540/2019 (PR). TP acknowledges ERC Advanced grant 694544-OMNES and ARRS research program P1-0402.
\end{acknowledgments}	
	
\bibliography{\bibpath/SYK,\bibpath/bibfile}

\let\addcontentsline\oldaddcontentsline% Restore \addcontentsline

\onecolumngrid

%SUPPLEMENTAL MATERIAL

\clearpage

\setcounter{table}{0}
\renewcommand{\thetable}{S\arabic{table}}%
\setcounter{figure}{0}
\renewcommand{\thefigure}{SM\arabic{figure}}%
\setcounter{equation}{0}
\renewcommand{\theequation}{S\arabic{equation}}%
\setcounter{page}{1}
\renewcommand{\thepage}{SM-\arabic{page}}%
\setcounter{secnumdepth}{3}
\setcounter{section}{0}
\renewcommand{\thesection}{\arabic{section}}%
\setcounter{subsection}{0}
\renewcommand{\thesubsection}{\arabic{section}.\arabic{subsection}}%

\begin{center}\Large{
		\textit{Supplemental Material for}\\
		\textbf{\MyTitle}\\
		by Lucas S\'a, Pedro Ribeiro, and Toma\v{z} Prosen
	}
\end{center}

{
\hypersetup{linkcolor=black}
\tableofcontents
}

\section*{Overview}
\addcontentsline{toc}{section}{Overview}

This Supplemental Material contains six Appendices. 

In Appendix~\ref{app:Lindbladian_path_integral}, we briefly review the derivation of the Keldysh path integral for the unitary evolution of bosons, complex fermions, and Majorana fermions. We then follow the same procedure to directly obtain the Keldysh path integral from the Lindblad equation.

In Appendix~\ref{app:micro_derivation_Lindblad_Keldysh}, we provide an alternative derivation of the Lindbladian path integral, starting from a microscopic unitary theory for the system plus environment. Averaging over the system-reservoir coupling to one-loop allows us to treat both Markovian and non-Markovian time evolution in the Keldysh formalism. We then specialize to the Markovian case to recover the results for the Lindbladian path integral. This procedure also allows us to provide a microscopic derivation of the memory kernel $K(z,z')$ that is consistent with causality. Finally, we also study the case of an SYK system coupled to an SYK thermal bath, which leads to a manifestly non-Markovian kernel.

In Appendix~\ref{app:details_effective_action}, we fill in the details of the derivation of the mean-field action for the collective fields $(G,\Sigma)$, Eq.~(\ref{eq:C_S_eff}), outlined in the Main Text. In particular, we comment on the choice of scaling of the couplings.

In Appendix~\ref{app:numerics}, we discuss in detail the solution of the Schwinger-Dyson equations. We start by defining the different components of the mean-field Green's functions and the relations between them, focusing, in particular, on the case of Majorana fermions. We then discuss in more detail the Keldysh rotation performed in the Main Text and derive Eqs.~(\ref{eq:SDE_sigma+-}) and (\ref{eq:SDE_rho+-}) (the Schwinger-Dyson equations for $\rho^\pm$ and $\sigma^\pm$). Then, we elaborate on the numerical method used to solve these equations. We also present numerical results for thermal equilibrium. Finally, we discuss the regime of validity of the self-consistent Lorentzian ansatz for the spectral function, Eq.~(\ref{eq:rho-_ansatz_largegamma}).

In Appendix~\ref{app:generalizations}, we present the generalization of the four-body SYK Lindbladian of the Main Text for higher $q$-body interactions.

In Appendix~\ref{app:ed}, we present some exact diagonalization results for finite $N$ and $M$. We study the spectral statistics, finding excellent agreement with random matrix theory and, therefore, quantum chaotic dynamics, and also compare the full spectrum to that of dense random Lindbladians.

\section{The Lindbladian path integral}
\label{app:Lindbladian_path_integral}

In this Appendix, we briefly review the derivation of the Keldysh path integral for the unitary evolution of bosons, complex fermions, and Majorana fermions. We then follow the same procedure to directly obtain the Keldysh path integral from the Lindblad equation.

\subsection{The Keldysh path integral}

The Keldysh generating function is defined as~\cite{kamenev2011}
\begin{align}
Z_{t_{\t f}} & 
=\Tr\left[\rho_{t_\t{f}}\right]
=\Tr\left[U\left(0,t_{\t f}\right)U\left(t_{\t f},0\right)\rho_{0}\right],
\end{align}
where $\rho_0$ and $\rho_{t_\t f}$ are the initial and final density matrices, $U\left(t,t'\right)=T_t\, e^{-\i\int_{t'}^{t}d\tau H(\tau)}$ for $t>t'$ ($T_t$ the time-ordering operator), and $U\left(t,t'\right)=U\left(t',t\right)^{\dagger}$ for $t<t'$. As it stands, $Z_{t_{\t f}}=1$, however, we will assume that the forward, $U\left(t_{\t f},0\right)$, and backward, $U\left(0,t_{\t f}\right)$, propagation can, in principle, be different. In practice, we can consider different source terms in each branch and vary the generating function with respect to the sources to obtain correlation functions of observables. 

Using a set of complex or Grassmann variables, $\xi^\pm(t)$, for bosons or fermions propapagating forward ($+$) or backward ($-$) in time and the associated coherent states, $\ket{\xi^\pm_t}$, the partition function reads as~\cite{kamenev2011}
\begin{align}
Z_{t_{\t f}} & =\int \sD\xi\ \tilde{\rho}_{0}
\exp\left\{\i\int_{0}^{t_{\t f}}\d t\left[ \bar{\xi}^{+}(t)\,\i\pd_{t}\xi^{+}(t)-\bar{\xi}^{-}(t)\,\i\pd_{t}\xi^{-}(t)-H^{+}(t)+H^{-}(t)\right] \right\},
\end{align}
where 
\begin{align}\label{eq:SM_H_trot}
\tilde{\rho}_{0}=\frac{\bra{\xi_{0}^{+}}\rho_{0}\ket{\zeta\xi_{0}^{-}}}{\sqrt{\braket{\xi_{0}^{-}}{\xi_{0}^{-}}\braket{\xi_{0}^{+}}{\xi_{0}^{+}}}},
\quad 
H^{+}(t)=\frac{\bra{\xi_{t+\d t}^{+}}H\ket{\xi_{t}^{+}}}{\braket{\xi_{t+\d t}^{+}}{\xi_{t}^{+}}},
\quad \text{and} \quad H^{-}(t)=\frac{\bra{\xi_{t}^{-}}H\ket{\xi_{t+\d t}^{-}}}{\braket{\xi_{t}^{-}}{\xi_{t+\d t}^{-}}}.
\end{align}
with $\zeta=\pm1$ for bosons or fermions, and the continuum limit $\d t\to 0$ is understood.

Allowing for more than one bosonic/fermionic field $\xi^\pm_i(t)$ and defining the Keldysh contour of integration, $\sC=\sC^+\cup \sC^- $, with $\d z=\pm \d t$
and $\xi_{i}\left(z\right)=\xi_{i}^{\pm}\left(t\right)$ for $z\in\mathcal{C^{\pm}}$,
we get 
\begin{align}\label{eq:SM_Keldysh_path_integral}
Z_{t_{\t f}}= & \int \sD\xi\ \tilde{\rho}_{0}\exp{\i\int_{\mathcal{C}}\d z\left[ \sum_i\bar{\xi}_{i}(z)\,\i\pd_{z}\xi_{i}(z)-H(z)\right] }.
\end{align}

\subsection{Majorana fermions}

Consider $2N$ Majorana operators, defined as
\begin{align}
\begin{split}
\chi_{2i}  &=\frac{1}{\sqrt{2}}\(c_{i}+c_{i}^{\dagger}\),\\
\chi_{2i-1} &=\frac{\i}{\sqrt{2}}\left(c_{i}-c_{i}^{\dagger}\right),
\end{split}
\end{align}
where $c^\dagger_i$ and $c_i$ are canonical creation and annihilation operators, that satisfy the Clifford algebra
\begin{align}
\left\{\chi_{i},\chi_{j}\right\}  & =\delta_{ij}.
\end{align}
We can then define the real Grassmann quantities $a_i$,
\begin{align}
\left(\begin{array}{c}
\xi_{i}\\
\bar{\xi}_{i}
\end{array}\right)=
& \frac{1}{\sqrt{2}}\left(\begin{array}{c}
a_{2i}-\i a_{2i+1}\\
a_{2i}+\i a_{2i+1}
\end{array}\right)=
\frac{1}{\sqrt{2}}\left(
\begin{array}{cc}
1 & -\i\\
1 & \i
\end{array}\right)
\left(\begin{array}{c}
a_{2i}\\
a_{2i+1}
\end{array}\right),
\end{align}
in terms of which the kinetic term of the action reads as
\begin{align}
\int_{\mathcal{C}}\d z\ \bar{\xi}_{i}(z)\,\i\pd_{z}\xi_{i}(z) & =\int_{\mathcal{C}}\d z\ \frac{1}{2}\left(\begin{array}{c}
\xi_{i}(z)\\
\bar{\xi}_{i}(z)
\end{array}\right)^{\dagger}\i\pd_{z}\left(\begin{array}{c}
\xi_{i}(z)\\
\bar{\xi}_{i}(z)
\end{array}\right)=\int_{\mathcal{C}}\d z\frac{1}{2}a_{i}(z)\,\i\pd_{z}a_{i}(z).
\end{align}
The Hamiltonian term is simply defined from the complex case, Eq.~(\ref{eq:SM_H_trot}).

\subsection{Lindbladian time evolution}

We proceed similarly to the standard Hamiltonian case. We start from the Lindblad equation 
\begin{align}
\pd_{t}\rho=\mathcal{\mathcal{L}}\left(\rho\right) 
& = -\i\left[H,\rho\right]
+\sum_{m}\(
2L_{m}\rho L_{m}^{\dagger}
-L_{m}^{\dagger}L_{m}\rho
-\rho L_{m}^{\dagger}L_{m}
\).
\end{align}
From its differential form,
\begin{align}
\rho(t+\d t) & =\rho(t)+\left(
-\i H\rho(t) + \i\rho(t)H
+\sum_{m}2L_{m}\rho(t)L_{m}^{\dagger}
-L_{m}^{\dagger}L_{m}\rho(t)
-\rho(t)L_{m}^{\dagger}L_{m}\right) \d t,
\end{align}
it is possible to write the matrix element $\bra{\xi_{t+\d t}^{+}}\rho(t+\d t)\ket{\xi_{t+\d t}^{-}}$
as a function of $\bra{\xi_{t}^{+}}\rho(t)\ket{\xi_{t}^{-}}$ using
the partition of the identity of coherent states  
\begin{equation}
\begin{split}
\bra{\xi_{t+\d t}^{+}}&\rho(t+\d t)\ket{\xi_{t+\d t}^{-}} =\int \sD\xi_{t}\bra{\xi_{t}^{+}}\rho(t)\ket{\xi_{t}^{-}}\\
&\times\exp{\left[
	\frac{\bar{\xi}_{t+\d t}^{+}-\bar{\xi}_{t}^{+}}{\d t}
	\xi_{t}^{+}
	+\bar{\xi}_{t}^{-}\frac{\xi_{t+\d t}^{-}-\xi_{t}^{-}}{\d t}
	-\i H^{+}(t)+\i H^{-}(t)
	+\sum_{m}\(2L_{m}^{+}(t)\bar{L}_{m}^{-}(t)
	-\Gamma^{+}(t)-\Gamma^{-}(t)
	\)\right] \d t},
\end{split}
\end{equation}
where we defined
\begin{align}
&L_{m}^{+}(t)=\frac{\bra{\xi_{t+\d t}^{+}}L_m\ket{\xi_{t}^{+}}}{\braket{\xi_{t+\d t}^{+}}{\xi_{t}^{+}}},
\quad
\bar{L}_{m}^{-}(t)=\frac{\bra{\xi_{t}^{-}}L_m^\dagger\ket{\xi_{t+\d t}^{-}}}{\braket{\xi_{t}^{-}}{\xi_{t+\d t}^{-}}},
\\
&\Gamma^+(t)=\frac{\bra{\xi_{t+\d t}^{+}}\sum_{m}L_{m}^{\dagger}L_{m}\ket{\xi_{t}^{+}}}{\braket{\xi_{t+\d t}^{+}}{\xi_{t}^{+}}},
\quad \text{and} \quad
\Gamma^-(t)=\frac{\bra{\xi_{t}^{-}}\sum_{m}L_{m}^{\dagger}L_{m}\ket{\xi_{t+\d t}^{-}}}{\braket{\xi_{t}^{-}}{\xi_{t+\d t}^{-}}}.
\end{align}
The generating function is given by iterating the procedure above, expressing $\rho_{t_{\t f}}$ in terms of $\rho_{0}$: 
\begin{align}
\begin{split}
Z_{t_{\t f}}  & =\Tr\left[\rho_{t_{\t f}}\right]=
\int \sD\xi\ 
\frac{\bra{\xi_{0}^{+}}\rho_{0}\ket{\zeta\xi_{0}^{-}}}{\braket{\xi_{0}^{+}}{\xi_{0}^{-}}}
\\
\times&\exp{
\sum_{t}\left[ 
\frac{\bar{\xi}_{t+\d t}^{+}-\bar{\xi}_{t}^{+}}{\d t}\xi_{t}^{+}
+\bar{\xi}_{t}^{-}\frac{\xi_{t+\d t}^{-}-\xi_{t}^{-}}{\d t}
-\i H^{+}(t)+\i H^{-}(t)
+\sum_{m}\(
2L_{m}^{+}(t)\bar{L}_{m}^{-}(t) -\Gamma^{+}(t)-\Gamma^{-}(t)
\) \right] \d t}.
\end{split}
\end{align}
In the continuum limit, we obtain
\begin{equation}
Z_{t_{\t f}}= \int \sD\xi\ \tilde{\rho}_0\,
\exp\left\{\i\int \d t\left[\bar{\xi}^{+}\,\i\pd_{t}\xi^{+}-\bar{\xi}^{-}\,\i\pd_{t}\xi^{-}-\left(H^{+}(t)-\i\Gamma^{+}(t)\right)+\left(H^{-}(t)+\i\Gamma^{-}(t)\right)-2\i \sum_m L_{m}^{+}(t)\bar{L}_{m}^{-}(t)\right]\right\}.
\end{equation}
To recover the expression obtained from the microscopic theory (see Appendix \ref{app:micro_derivation_Lindblad_Keldysh}), we simply make the change of variables $\xi_{t}^{-}\to\zeta\xi_{t}^{-}$
\begin{align}\label{eq:SM_Keldysh_Lindblad}
Z_{t_{\t f}}= & \int \sD\xi\ \tilde{\rho}_0\exp{\i\int \d t\left[\bar{\xi}^{+}\,\i\pd_{t}\xi^{+}-\bar{\xi}^{-}\,\i\pd_{t}\xi^{-}-\left(H^{+}(t)-\i\Gamma^{+}(t)\right)+\left(H^{-}(t)+\i\Gamma^{-}(t)\right)-2\i\zeta' \sum_m L_{m}^{+}(t)\bar{L}_{m}^{-}(t)\right]},
\end{align}
where $\zeta'=\pm1$ depending on the bosonic or fermionic nature of the jump operators $L_m$. Note that $\zeta$ and $\zeta'$ may differ, since, e.g., for a system of fermions ($\zeta=-1$), the jump operators may be bosonic ($\zeta'=1$). This is the case for the quadratic jump operators of the Main Text, where we set $\zeta'=1$ throughout.

\section{Microscopic derivation of the Lindbladian path integral}
\label{app:micro_derivation_Lindblad_Keldysh}

In this Appendix, we provide an alternative derivation of the Lindbladian path integral, starting from a microscopic unitary theory for the system plus environment. Averaging over the system-reservoir coupling to one-loop allows us to treat both Markovian and non-Markovian time evolution in the Keldysh formalism. We then specialize to the Markovian case to recover the results for the Lindbladian path integral. This procedure also allows us to provide a microscopic derivation of the memory kernel $K(z,z')$ that is consistent with causality. Finally, we also study the case of an SYK system coupled to an SYK thermal bath, which leads to a manifestly non-Markovian kernel.

\subsection{System-reservoir coupling to one loop}
\label{app:pertubation_theory}

Our starting point is the Keldsyh action~(\ref{eq:SM_Keldysh_path_integral}) for the Hamiltonian time evolution. Assuming a system, $\t S$, in contact with a macroscopic reservoir,
$\t R$, we can partition the degrees of freedom into $\t S$ and
$\t R$. We will further assume the Born approximation, i.e., the $\t{S-R}$ coupling can be treated in a one-loop approximation. Assuming a coupling
of the form 
\begin{align}\label{eq:coupling_SR}
H_{\t{SR}} & =\sum_{\mu}\hat{X}_{\mu}\hat{Y}_{\mu}^{\dagger},
\end{align}
where $\hat{X}_{\mu}$ and $\hat{Y}_{\mu}$ are operators in $\t S$ and $\t R$, respectively, that $\av{\hat{Y}_{\mu}}=0$, and that at
the initial time $\t S$ and $\t R$ are uncorrelated, we obtain 
\begin{equation}\label{eq:SM_Keldysh_general}
\begin{split}
Z_{t_{\t f}} & =\int \sD\xi\ \tilde{\rho}_{0}
\exp{\i\int_{\mathcal{C}}\d z\left\{ \bar{\xi}(z)\,\i\pd_{z}\xi(z)-H(z)\right\} -\int_{\mathcal{C}}\d z\d z'\sum_{\mu\mu'}\bar{X}_{\mu}(z)\av{T_{z}\hat{Y}_{\mu}(z)\hat{Y}_{\mu'}^{\dagger}(z')}X_{\mu'}(z')}\\
& =\int \sD\xi\ \tilde{\rho}_{0}
\exp{\i\int_{\mathcal{C}}\d z\left\{ \bar{\xi}(z)\,\i\pd_{z}\xi(z)-H(z)\right\} -\int_{\mathcal{C}}\d z\d z'\sum_{\mu\mu'}\bar{X}_{\mu}(z)\,\i\Omega_{\mu\mu'}\left(z,z'\right)X_{\mu'}(z')},
\end{split}
\end{equation}
where the integration is now done solely over the system's degrees of freedom,
\begin{align}\label{eq:Omega_def}
\Omega_{\mu\mu'}\left(z,z'\right) & =-\i\av{T_{z}\hat{Y}_{\mu}(z)\hat{Y}_{\mu'}^{\dagger}(z')}
\end{align}
are the contour-ordered correlation functions of the environment, and
\begin{align}
X_{\mu}(z)= & \frac{\bra{\xi_{z+\d z}}\hat{X}_{\mu}\ket{\xi_{z}}}{\braket{\xi_{z+\d z}}{\xi_{z}}},\\
\bar{X}_{\mu}(z)= & \frac{\bra{\xi_{z+\d z}}\hat{X}_{\mu}^{\dagger}\ket{\xi_{z}}}{\braket{\xi_{z+\d z}}{\xi_{z}}}.
\end{align}
Equation~(\ref{eq:SM_Keldysh_general}) is obtained to one-loop order in the system-reservoir coupling. It becomes exact in the case of Hamiltonians quadratic in the creation and annihilation operators of the environment with linear system-environment couplings $\hat{Y}_{\mu}$. 
For the Hamiltonian to be Hermitian it is required that
$\sum_{\mu}\hat{X}_{\mu}\hat{Y}_{\mu}^{\dagger}=\zeta'\sum_{\mu}\hat{X}_{\mu}^{\dagger}\hat{Y}_{\mu}$, where $\zeta'=\pm1$ depending on the bosonic or fermionic nature of the operators $\hat{Y}_{\mu}$ (equivalently $\hat{X}_{\mu}$), i.e., the sum over $\mu$ has to run over all operators $\hat{X}_{\mu}$
and their conjugates $\hat{X}_{\mu}^{\dagger}$. It is convenient to define the notation $\hat{X}_{\mu}^{\dagger}=\hat{X}_{\bar{\mu}}=\hat{X}_{-\mu}$,
such that $\sum_{\mu}\hat{X}_{\mu}\hat{Y}_{\mu}^{\dagger}=\sum_{\mu>0}\left(\hat{X}_{\mu}\hat{Y}_{\mu}^{\dagger}+\zeta'\hat{X}_{\bar{\mu}}^{\dagger}\hat{Y}_{\bar{\mu}}\right)$.
Therefore we can also write
\begin{align}
\label{eq:SM_XOmegaX1}
\sum_{\mu\mu'}\bar{X}_{\mu}(z)\,\Omega_{\mu\mu'}\left(z,z'\right)X_{\mu'}(z') & =-\i\sum_{\mu\mu'}\bar{X}_{\mu}(z)\av{T_{z}\hat{Y}_{\mu}(z)\hat{Y}_{\mu'}^{\dagger}(z')}X_{\mu'}(z')\\
\label{eq:SM_XOmegaX2}
& =-\i\sum_{\mu\mu'} X_{\mu'}(z')\av{T_{z}\hat{Y}_{\mu'}^{\dagger}(z')\hat{Y}_{\mu}(z)}\bar{X}_{\mu}(z)=\sum_{\mu\mu'}X_{\mu}(z)\,\Omega_{\bar{\mu}\bar{\mu}'}\left(z,z'\right)\bar{X}_{\mu'}(z')\\
\label{eq:SM_XOmegaX3}
& =-\i\zeta' \sum_{\mu\mu'}X_{\mu}(z)\av{T_{z}\hat{Y}_{\mu}^{\dagger}(z)\hat{Y}_{\mu'}^{\dagger}(z')}X_{\mu'}(z')=\zeta'\sum_{\mu\mu'} X_{\mu}(z)\,\Omega_{\bar{\mu}\mu'}\left(z,z'\right)X_{\mu'}(z')\\
\label{eq:SM_XOmegaX4}
& =-\i\sum_{\mu\mu'}\bar{X}_{\mu}(z)\av{T_{z}\hat{Y}_{\mu}(z)\hat{Y}_{\mu'}}\bar{X}_{\mu'}(z')=\zeta'\sum_{\mu\mu'} \bar{X}_{\mu}(z)\,\Omega_{\mu\bar{\mu}'}\left(z,z'\right)\bar{X}_{\mu'}(z'),
\end{align}
where we defined 
\begin{align}
\Omega_{\mu\mu'}\left(z,z'\right) & =-\i\av{T_z\hat{Y}_{\mu}(z)\hat{Y}_{\mu'}^{\dagger}(z')},\\
\Omega_{\bar{\mu}\mu'}\left(z,z'\right) & =-\i\av{T_z\hat{Y}_{\mu}^{\dagger}(z)\hat{Y}_{\mu'}^{\dagger}(z')},\\
\Omega_{\mu\bar{\mu}'}\left(z,z'\right) & =-\i\av{T_z\hat{Y}_{\mu}(z)\hat{Y}_{\mu'}(z')},\\
\Omega_{\bar{\mu}\bar{\mu}'}\left(z,z'\right) & =-\i\av{T_z\hat{Y}_{\mu}^{\dagger}(z)\hat{Y}_{\mu'}(z')}.
\end{align}

In terms of real time, $\d z=\pm \d t$ for the forward and backward contour, respectively, the S-R coupling term reads as
\begin{align}
\begin{split}
-\i\int_{\mathcal{C}}\d z\d z'\sum_{\mu\mu'} &
\bar{X}_{\mu}(z)\Omega_{\mu\mu'}\left(z,z'\right)X_{\mu'}(z')= 
-\i\int \d t\d t' \sum_{\mu\mu'}
\left[
\bar{X}_{\mu}^{+}(t)\,\Omega_{\mu\mu'}^{\rmT}\left(t,t'\right)X_{\mu'}^{+}(t')\right.\\
&\left.
+\bar{X}_{\mu}^{-}(t)\,\Omega_{\mu\mu'}^{\rmTb}\left(t,t'\right)X_{\mu'}^{-}(t')
-\bar{X}_{\mu}^{+}(t)\,\Omega_{\mu\mu'}^{<}\left(t,t'\right)X_{\mu'}^{-}(t')
-\bar{X}_{\mu}^{-}(t)\,\Omega_{\mu\mu'}^{>}\left(t,t'\right)X_{\mu'}^{+}(t')\right],
\end{split}
\end{align}
where we introduced the greater, lesser, time-ordered, and anti-time-ordered components of $\Omega(z,z')$:
\begin{align}
\label{eq:SM_Omega>}
\Omega_{\mu\mu'}^{>}\left(t,t'\right) & =-\i\av{\hat{Y}_{\mu}(t)\hat{Y}_{\mu'}^{\dagger}(t')},\\
\label{eq:SM_Omega<}
\Omega_{\mu\mu'}^{<}\left(t,t'\right) & =-\i\zeta' \av{\hat{Y}_{\mu'}^{\dagger}(t')\hat{Y}_{\mu}(t)},\\
\label{eq:SM_OmegaT}
\Omega_{\mu\mu'}^{\rmT}\left(t,t'\right) & =\Theta\left(t-t'\right)\Omega_{\mu\mu'}^{>}\left(t,t'\right)+\Theta\left(t'-t\right)\Omega_{\mu\mu'}^{<}\left(t,t'\right),\\
\label{eq:SM_OmegaTb}
\Omega_{\mu\mu'}^{\rmTb}\left(t,t'\right) & =\Theta\left(t'-t\right)\Omega_{\mu\mu'}^{>}\left(t,t'\right)+\Theta\left(t-t'\right)\Omega_{\mu\mu'}^{<}\left(t,t'\right).
\end{align}
We can rewrite the terms of the action that couple different branches of the Keldysh contour as
\begin{equation}\label{eq:SM_action_coupling_branches}
\begin{split}
&-\i\int \d t\d t'\sum_{\mu\mu'}\left[ -\bar{X}_{\mu}^{+}(t)\,\Omega_{\mu\mu'}^{<}\left(t,t'\right)X_{\mu'}^{-}(t')-\bar{X}_{\mu}^{-}(t)\,\Omega_{\mu\mu'}^{>}\left(t,t'\right)X_{\mu'}^{+}(t')\right]
\\
=&-\i\int \d t\d t'\sum_{\mu\mu'}\left[ -\bar{X}_{\mu}^{+}(t)\,\Omega_{\mu\mu'}^{<}\left(t,t'\right)X_{\mu'}^{-}(t')-\zeta'\bar{X}_{\mu}^{+}(t)\,\Omega_{\bar{\mu}'\bar{\mu}}^{>}\left(t',t\right)X_{\mu'}^{-}(t')\right] \\
=&\int \d t\d t'\sum_{\mu\mu'}\bar{X}_{\mu}^{+}(t)\left[2\i\Omega_{\mu\mu'}^{<}\left(t,t'\right)\right]X_{\mu'}^{-}(t'),
\end{split}
\end{equation}
where the last equality follows from the identity
\begin{align}
\Omega_{\bar{\mu}'\bar{\mu}}^{>}\left(t',t\right) & =-\i\av{\hat{Y}_{\mu}^{\dagger}(t)\hat{Y}_{\mu'}(t')}=\zeta'\Omega_{\mu\mu'}^{<}\left(t,t'\right),
\end{align}
while we also have
\begin{align}\label{eq:SM_hermiticity_Omega<}
\Omega_{\mu'\mu}^{<}\left(t',t\right)^{*} & =i\zeta'\av{\hat{Y}_{\mu'}^{\dagger}(t')\hat{Y}_{\mu}(t)}=-\Omega_{\mu\mu'}^{<}\left(t,t'\right).
\end{align}

We conclude that the Keldysh path integral for the (generally non-Markovian) dynamics of the reduced system is given by:
\begin{equation}\label{eq:SM_non_Markovian_action}
\begin{split}
Z_{t_{\t f}} & =\int \sD\xi\ \tilde{\rho}_{0}
\exp\left\{\i\int \d t\left[\bar{\xi}^{+}(t)\,\i\pd_{t}\xi^{+}(t)-\bar{\xi}^{-}(t)\,\i\pd_{t}\xi^{-}(t)-H^+(t)+H^-(t)\right]\right.
\\
&\left.-\int \d t\d t'\sum_{\mu\mu' } \( \bar{X}_{\mu}^{+}(t)\,\i\Omega_{\mu\mu'}^{\rmT}\left(t,t'\right)X_{\mu'}^{+}(t')+\bar{X}_{\mu}^{-}(t)\,\i\Omega_{\mu\mu'}^{\rmTb}\left(t,t'\right)X_{\mu'}^{-}(t')-\bar{X}_{\mu}^{+}(t)\,2\i\Omega_{\mu\mu'}^{<}\left(t,t'\right)X_{\mu'}^{-}(t')\) \right\}.
\end{split}
\end{equation}

\subsection{Markovian approximation}

Assuming that the time-scales of the system are much smaller than those of the reservoir, we can approximate the correlation functions by an equal-time correlation function:
\begin{align}
\label{eq:SM_Markovian_Omega<}
\i\Omega_{\mu\mu'}^{<}\left(t,t'\right) & \simeq\zeta'\delta\left(t-t'\right)M_{\mu\mu'}\left(t\right),
\\
\label{eq:SM_Markovian_Omega>}
\i\Omega_{\mu\mu'}^{>}\left(t,t'\right) & =\i\zeta'\Omega_{\bar{\mu}'\bar{\mu}}^{<}\left(t',t\right)\simeq\delta\left(t-t'\right)M_{\bar{\mu}'\bar{\mu}}\left(t\right).
\end{align}
By Hermiticity of $\i\Omega_{\mu\mu'}^{<}\left(t,t'\right)$, Eq.~(\ref{eq:SM_hermiticity_Omega<}), we have
that $M^{\dagger}=M$. Moreover, the positive-definiteness of $\sum_m L_m^\dagger L_m$ implies that $M$ is also positive-definite. (This can easily be seen in the time-independent case as follows. By inverting Eq.~(\ref{eq:SM_Markovian_Omega<}), we can write
\begin{align}
M_{\mu\mu'} = \int \d t \d t' \ \i\Omega_{\mu\mu'}^{<}\left(t,t'\right)
= \av{
 \int \d t'\, \hat{Y}_{t'\mu'}^{\dagger}
 \int \d t\, \hat{Y}_{t\mu}
 }.
\end{align}
The last term is the average of a positive-definite operator and therefore is non-negative.)
We can then decompose
\begin{equation}\label{eq:SM_M_def} M_{\mu\mu'}\left(t\right)=\sum_{m}v_{\mu}^{m}(t)\lambda_{m}v_{\mu'}^{m*}(t),
\end{equation}
with $\lambda_{m}\ge0$. For the term in the action coupling different branches, Eq.~(\ref{eq:SM_action_coupling_branches}), this implies 
\begin{align}
\int \d t \d t' \sum_{\mu\mu' }
\bar{X}_{\mu}^{+}(t)\left[2\i\Omega_{\mu\mu'}^{<}\left(t,t'\right)\right]X_{\mu'}^{-}(t') 
& =\int \d t\sum_{\mu\mu'}\bar{X}_{\mu}^{+}(t)\left[2\zeta' M_{\mu\mu'}\left(t\right)\right]X_{\mu'}^{-}(t)\\
& =\int \d t\sum_{m}2\zeta'\left[\bar{X}_{\mu}^{+}(t)v_{\mu}^{m}(t)\sqrt{\lambda_{m}}\right]\left[\sqrt{\lambda_{m}}v_{\mu'}^{m*}(t)X_{\mu'}^{-}(t)\right]\\
& =\int \d t\ \sum_{m}2\zeta' L_{m}^{+}(t)\bar{L}_{m}^{-}(t),
\end{align}
where we defined the jump operators
\begin{align}\label{eq:SM_jump_op_def}
L_{m}(t) & =\sum_\mu\hat{X}_{\mu}^{\dagger}v_{\mu}^{m}(t)\sqrt{\lambda_{m}}
\end{align}
and their contour representation
\begin{align}
\label{eq:SM_jump_op_contour+}
L_{m}^{+}(t) & =\sum_\mu \bar{X}_{\mu}^{+}(t)v_{\mu}^{m}(t)\sqrt{\lambda_{m}}=\frac{\bra{\xi_{t+\d t}^{+}}\sum_\mu\hat{X}_{\mu}^{\dagger}v_{\mu}^{m}(t)\sqrt{\lambda_{m}}\ket{\xi_{t}^{+}}}{\braket{\xi_{t+\d t}^{+}}{\xi_{t}^{+}}},\\
\label{eq:SM_jump_op_contour-}
\bar{L}_{m}^{-}(t) & =   \sum_\mu \sqrt{\lambda_{m}}v_{\mu'}^{m*}(t)X_{\mu'}^{-}(t)=\frac{\bra{\xi_{t}^{-}}\sum_\mu\sqrt{\lambda_m}v_{\mu'}^{m*}(t)\hat{X}_{\mu'}\ket{\xi_{t+\d t}^{-}}}{\braket{\xi_{t}^{-}}{\xi_{t+\d t}^{-}}}.
\end{align}

The terms acting within a single branch of the contour are much more delicate to deal with because of time-ordering.
We use the intuition from what we expect for the Lindblad case to postulate the following equal-time limits
\begin{align}
\label{eq:SM_limit1}
\lim_{t\to t'}\bar{X}_{\mu}^{+}(t)X_{\mu'}^{+}(t')\,\Theta\left(t-t'\right) & =\frac{\bra{\xi_{t+\d t}^{+}}\hat{X}_{\mu}^{\dagger}\hat{X}_{\mu'}\ket{\xi_{t}^{+}}}{\braket{\xi_{t+\d t}^{+}}{\xi_{t}^{+}}}\,\Theta\left(t-t'\right),
\\
\label{eq:SM_limit2}
\lim_{t\to t'}\bar{X}_{\mu}^{+}(t)X_{\mu'}^{+}(t')\,\Theta\left(t'-t\right) & =\zeta'\frac{\bra{\xi_{t+\d t}^{+}}\hat{X}_{\mu'}\hat{X}_{\mu}^{\dagger}\ket{\xi_{t}^{+}}}{\braket{\xi_{t+\d t}^{+}}{\xi_{t}^{+}}}\,\Theta\left(t-t'\right),
\\
\label{eq:SM_limit3}
\lim_{t\to t'}\bar{X}_{\mu}^{-}(t)X_{\mu'}^{-}(t')\,\Theta\left(t-t'\right) & =\zeta'\frac{\bra{\xi_{t}^{-}}\hat{X}_{\mu'}\hat{X}_{\mu}^{\dagger}\ket{\xi_{t+\d t}^{-}}}{\braket{\xi_{t}^{-}}{\xi_{t+\d t}^{-}}}\,\Theta\left(t-t'\right),
\\
\label{eq:SM_limit4}
\lim_{t\to t'}\bar{X}_{\mu}^{-}(t)X_{\mu'}^{-}(t')\,\Theta\left(t'-t\right) & =\frac{\bra{\xi_{t}^{-}}\hat{X}_{\mu}^{\dagger}\hat{X}_{\mu'}\ket{\xi_{t+\d t}^{-}}}{\braket{\xi_{t}^{-}}{\xi_{t+\d t}^{-}}}\,\Theta\left(t-t'\right).
\end{align}
For the forward branch, we obtain
\begin{equation}\label{eq:SM_forward_action_1}
\begin{split}
&\int \d t \d t'\sum_{\mu\mu' }
\bar{X}_{\mu}^{+}(t)\left[\i\Omega_{\mu\mu'}^{\rmT}\left(t,t'\right)\right]X_{\mu'}^{+}(t')
\\
=&\int \d t\d t'\sum_{\mu\mu' }\left\{ \bar{X}_{\mu}^{+}(t)\left[\i\Omega_{\mu\mu'}^{>}\left(t,t'\right)\Theta\left(t-t'\right)\right]X_{\mu'}^{+}(t')+\bar{X}_{\mu}^{+}(t)\left[\i\Omega_{\mu\mu'}^{<}\left(t,t'\right)\Theta\left(t'-t\right)\right]X_{\mu'}^{+}(t')\right\}
\\
=&\int \d t\d t'\sum_{\mu\mu' }\left\{ \frac{\bra{\xi_{t+\d t}^{+}}\hat{X}_{\mu}^{\dagger}\hat{X}_{\mu'}\ket{\xi_{t}^{+}}}{\braket{\xi_{t+\d t}^{+}}{\xi_{t}^{+}}}\left[\i\Omega_{\mu\mu'}^{>}\left(t,t'\right)\Theta\left(t-t'\right)\right]+\zeta'\frac{\bra{\xi_{t+\d t}^{+}}\hat{X}_{\mu'}\hat{X}_{\mu}^{\dagger}\ket{\xi_{t}^{+}}}{\braket{\xi_{t+\d t}^{+}}{\xi_{t}^{+}}}\left[\i\Omega_{\mu\mu'}^{<}\left(t,t'\right)\Theta\left(t'-t\right)\right]\right\},
\end{split}
\end{equation}
where we used the definition of the time-ordered and anti-time-ordered correlation functions, Eqs.~(\ref{eq:SM_OmegaT}) and (\ref{eq:SM_OmegaTb}), and the limits of Eqs.~(\ref{eq:SM_limit1}) and (\ref{eq:SM_limit2}) to obtain the first and second equalities, respectively.
By renaming indices and recalling that $\hat{X}_{\bar{\mu}}=\hat{X}^\dagger_{\mu}$, the second term in the last line of Eq.~(\ref{eq:SM_forward_action_1}) can be rewritten as
\begin{equation}\label{eq:SM_forward_action_2}
\frac{\bra{\xi_{t+\d t}^{+}}\hat{X}_{\mu'}\hat{X}_{\mu}^{\dagger}\ket{\xi_{t}^{+}}}{\braket{\xi_{t+\d t}^{+}}{\xi_{t}^{+}}}\i\Omega_{\mu\mu'}^{<}\left(t,t'\right)
=
\frac{\bra{\xi_{t+\d t}^{+}}\hat{X}_{\mu}^{\dagger}\hat{X}_{\mu'}\ket{\xi_{t}^{+}}}{\braket{\xi_{t+\d t}^{+}}{\xi_{t}^{+}}}\i\Omega_{\bar{\mu}'\bar{\mu}}^{<}\left(t,t'\right).
\end{equation}
Plugging Eq.~(\ref{eq:SM_forward_action_2}) into Eq.~(\ref{eq:SM_forward_action_1}) and using the Markovian approximation of Eqs.~(\ref{eq:SM_Markovian_Omega<}) and (\ref{eq:SM_Markovian_Omega>}), we obtain
\begin{equation}\label{eq:SM_forward_action_3}
\int \d t \d t' \sum_{\mu\mu' }
\bar{X}_{\mu}^{+}(t)\left[\i\Omega_{\mu\mu'}^{\rmT}\left(t,t'\right)\right]X_{\mu'}^{+}(t')
=\int \d t\d t'\sum_{\mu\mu'}\frac{\bra{\xi_{t+\d t}^{+}}\hat{X}_{\mu}^\dagger\hat{X}_{\mu'}\ket{\xi_{t}^{+}}}{\braket{\xi_{t+\d t}^{+}}{\xi_{t}^{+}}}M_{\bar{\mu}'\bar{\mu}}\left(t\right)\delta\left(t-t'\right).
\end{equation}
We can further use Eq.~(\ref{eq:SM_XOmegaX2}) to rewrite Eq.~(\ref{eq:SM_forward_action_3}) as
\begin{equation}\label{eq:SM_forward_action_4}
\int \d t \d t' \sum_{\mu\mu'}
\bar{X}_{\mu}^{+}(t)\,\i\Omega_{\mu\mu'}^{\rmT}\left(t,t'\right) X_{\mu'}^{+}(t')
=\int \d t\d t'\sum_{\mu\mu'}\frac{\bra{\xi_{t+\d t}^{+}}\hat{X}_{\mu}\hat{X}_{\mu'}^\dagger\ket{\xi_{t}^{+}}}{\braket{\xi_{t+\d t}^{+}}{\xi_{t}^{+}}}M_{\mu'\mu}\left(t\right)\delta\left(t-t'\right),
\end{equation}
and, finally, use Eqs.~(\ref{eq:SM_M_def}) and (\ref{eq:SM_jump_op_def}) to arrive at
\begin{equation}
\begin{split}
\int \d t \d t' \sum_{\mu\mu'}
\bar{X}_{\mu}^{+}(t)\,\i\Omega_{\mu\mu'}^{\rmT}\left(t,t'\right)X_{\mu'}^{+}(t')
=&\int \d t\ \frac{\bra{\xi_{t+\d t}^{+}}\sum_{m}\left(\sum_\mu\hat{X}_{\mu}v_{\mu}^{m*}(t)\sqrt{\lambda_{m}}\right)\left(\sum_{\mu'}\sqrt{\lambda_{m}}v_{\mu'}^{m}(t)\hat{X}_{\mu'}^{\dagger}\right)\ket{\xi_{t}^{+}}}{\braket{\xi_{t+\d t}^{+}}{\xi_{t}^{+}}}
\\
=&\int \d t\ \frac{\bra{\xi_{t+\d t}^{+}}\sum_{m}L_{m}^{\dagger}(t)L_{m}(t)\ket{\xi_{t}^{+}}}{\braket{\xi_{t+\d t}^{+}}{\xi_{t}^{+}}}
=\int \d t\ \Gamma^{+}(t).
\end{split}
\end{equation}
We proceed similarly for the backward branch:
\begin{align}
\int \d t \d t'\sum_{\mu\mu'}\bar{X}_{\mu}^{-}(t)\,\i\Omega_{\mu\mu'}^{\rmTb}\left(t,t'\right)X_{\mu'}^{-}(t') & =\int \d t\ \frac{\bra{\xi_{t}^{-}}\sum_{m}L_{m}^{\dagger}(t)L_{m}(t)\ket{\xi_{t+\d t}^{-}}}{\braket{\xi_{t}^{-}}{\xi_{t+\d t}^{-}}}=\int \d t\ \Gamma^{-}(t).
\end{align}

Finally, the partition function in the Markovian limit becomes 
\begin{align}\label{eq:SM_Z_micro_final}
Z_{t_{\t f}} & =\int \sD\xi\ \tilde{\rho}_{0}
\exp{\i\int \d t\left[\bar{\xi}^{+}\,\i\pd_{t}\xi^{+}-\bar{\xi}^{-}\,\i\pd_{t}\xi^{-}-\left(H^{+}(t)-\i\Gamma^{+}(t)\right)+\left(H^{-}(t)+\i\Gamma^{-}(t)\right)-2\i\zeta'\sum_m L_{m}^{+}(t)\bar{L}_{m}^{-}(t)\right]},
\end{align}
which coincides with the result obtained directly from the Lindblad equation in Appendix~\ref{app:Lindbladian_path_integral}, Eq.~(\ref{eq:SM_Keldysh_Lindblad}).

\subsection{The memory kernel}
To make contact with the expression for the action written down in the Main Text, Eq.~(\ref{eq:C_Keldysh_action}), it remains to derive the explicit form of the memory kernel $K(z,z')$ from Eq.~(\ref{eq:SM_Z_micro_final}), namely,
\begin{align}
\label{eq:SM_K_kernel_T}
K^\rmT(t,t')&=K(t^+,t'^+)=\zeta'\dirac{t-t'},
\\
\label{eq:SM_K_kernel_aT}
K^\rmTb(t,t')&=K(t^-,t'^-)=\zeta'\dirac{t-t'},
\\
\label{eq:SM_K_kernel_<}
K^<(t,t')&=K(t^+,t'^-)=2\zeta'\dirac{t-t'},
\\
\label{eq:SM_K_kernel_>}
K^>(t,t')&=K(t^-,t'^+)=0.
\end{align}
Note there is a relative minus sign for the backward branch $\sC^-$, $\d z=\pm\d t$ for $z\in\sC^\pm$, that leads to all components of the kernels having the same sign. Moreover, there is an extra factor of $\zeta'$ in Eq.~(\ref{eq:SM_K_kernel_<}) compared to Eq.~(\ref{eq:K_kernel_<}), which arises because we set $\zeta'=1$ throughout the Main Text, given the bosonic nature of the quadratic SYK jump operators.

To identify the correlation functions of the environment, $\Omega_{\mu\mu'}\left(z,z'\right)$, we have to relate the Markovian action (\ref{eq:SM_Keldysh_Lindblad}) with the non-Markovian one, Eq.~(\ref{eq:SM_Keldysh_general}), essentially running the procedure of the previous section backwards. However, such a procedure is not unique. The simplest choice is to identify the jump operators with the system operators $\hat{X}_{\mu}$ themselves. This amounts to restricting ourselves to microscopic kernels $\Omega_{\mu\mu'}$ without inter-channel coupling (no anomalous terms in the bath Hamiltonian), i.e., that satisfy 
\begin{equation}\label{eq:Omega_restricted}
\Omega_{\mu\mu'}(z,z')=
\delta_{\mu\mu'}\Omegatilde_\mu(z,z').
\end{equation}
More concretely, if we choose
\begin{align}
v_{\mu}^{m}(t)\sqrt{\lambda_{m}} & =\delta_{\mu m},
\end{align}
then Eq.~(\ref{eq:SM_jump_op_contour+}) gives
\begin{equation}\label{eq:SM_simple_jump_ops}
\begin{cases}
\bar{X}_{\mu=m}^{+}(t) =L_{m}^{+}(t), & \text{for\ }\mu>0,\\
\bar{X}_{\mu=-m}^{+}(t)=X_{\mu=m}^{+}(t)=\bar{L}_{m}^{+}(t), &\text{for\ }\mu<0,
\end{cases}
\end{equation}
and similarly for $L_{m}^-(t)$. In this way, we get
\begin{align}
M_{\mu\mu'} & =\sum_{m}v_{\mu}^{m}(t)\lambda_{m}v_{\mu'}^{m*}(t)=\delta_{\mu\mu'}\Theta_{\mu},
\end{align}
with
\begin{align}
\Theta_{\mu} & =\begin{cases}
1, & \mu>0\\
0, & \mu<0
\end{cases}.
\end{align}
Therefore,
\begin{align}
\label{eq:Omega_tilde_components<}
& \i \Omegatilde_\mu^<(t,t')=\zeta'\, f(t,t')\, \Theta_\mu,
\\
\label{eq:Omega_tilde_components>}
& \i \Omegatilde_\mu^>(t,t')=\zeta'\, \i \Omegatilde^<_{\bar{\mu}}(t',t)=f(t',t)\, \Theta_{\bar{\mu}}.
\end{align}
where for the moment we have allowed for a non-Markovian time kernel $f(t,t')$. (We note that the $f(t,t')$ must be independent of the channel index $\mu$ for us to be able to do the averages over $\Gamma_{ijkl}$ in the Lindbladian SYK model. At most, we can introduce a channel-dependent multiplicative constant that would weight the variance of each channel in $\Gamma_{ijkl}$.)

We now focus on the dissipative contribution to the general Keldysh action, Eq.~(\ref{eq:SM_Keldysh_general}),
\begin{equation}\label{eq:action_micro_X}
-\i \int_\sC \d z \d z'\sum_{\mu\mu'} \conj{X}_\mu(z)\Omega_{\mu\mu'}(z,z')X_{\mu'}(z').
\end{equation}
We expand the sum over (negative and positive) $\mu$ in Eq.~(\ref{eq:action_micro_X}) as sums over (positive only) $m$:
\begin{equation}
\begin{split}
\sum_{\mu\mu'} \conj{X}_\mu(z)\Omega_{\mu\mu'}(z,z')X_{\mu'}(z')
=\sum_{m,m'=1}^M &\( \conj{X}_m(z)\Omega_{mm'}(z,z')X_{m'}(z')
+\conj{X}_m(z)\Omega_{m,-m'}(z,z')X_{-m'}(z')
\right.\\
&\left.+
\conj{X}_{-m}(z)\Omega_{-mm'}(z,z')X_{m'}(z') +\conj{X}_{-m}(z)\Omega_{-m,-m'}(z,z')X_{-m'}(z')
\).
\end{split}
\end{equation}
Because of the constraint (\ref{eq:Omega_restricted}), we have $\Omega_{m,-m'}=\Omega_{-m,m'}=0$, and using Eq.~(\ref{eq:SM_simple_jump_ops}), the dissipative contribution to Eq.~(\ref{eq:action_micro_X}) reads as
\begin{equation}
\begin{split}
-\i \int_\sC \d z \d z'\sum_{\mu\mu'} \conj{X}_\mu(z)\Omega_{\mu\mu'}(z,z')X_{\mu'}(z')
&=-\int_\sC \d z \d z'\sum_m\left[
L_m(z)\,\i\Omegatilde_m(z,z')\conj{L}_m(z')
+\conj{L}_m(z)\,\i\Omegatilde_{-m}(z,z')L_m(z')
\right]
\\
&= -\int_\sC \d z \d z'\sum_m \, L_m(z) K(z,z')\conj{L}_m(z'),
\end{split}
\end{equation}
where we defined the memory kernel
\begin{equation}\label{eq:micro_def_K}
K(z,z')=\i\Omegatilde_m(z,z')+\zeta' \i\Omegatilde_{-m}(z',z).
\end{equation}
Using Eqs.~(\ref{eq:Omega_tilde_components<}) and (\ref{eq:Omega_tilde_components>}), we can compute the lesser ($(z,z')\in \sC^+\times\sC^-$),
\begin{equation}
K^<(t,t')
=\i\Omegatilde^<_m(t,t')
+\zeta' \i\Omegatilde^>_{-m}(t',t)
=2\i \Omegatilde_m^<(t,t')=2\zeta' f(t,t'),
\end{equation}
greater ($(z,z')\in \sC^-\times\sC^+$),
\begin{equation}
K^>(t,t')
=\i\Omegatilde^>_m(t,t')
+\zeta' \i\Omegatilde^<_{-m}(t',t)
=0,
\end{equation}
time-ordered ($(z,z')\in \sC^+\times\sC^+$),
\begin{equation}
\begin{split}
K^\rmT(t,t')
&=\i\Omegatilde^\rmT_m(t,t')
+\zeta' \i\Omegatilde^\rmT_{-m}(t',t)
\\
&=\heav{t-t'}\i\Omegatilde_m^>(t,t')
+\heav{t'-t}\i\Omegatilde_m^<(t,t')
\zeta'\heav{t'-t}\i\Omegatilde_{-m}^>(t',t)
+\zeta'\heav{t-t'}\i\Omegatilde_{-m}^<(t',t)
\\
&=2\heav{t'-t}\i\Omegatilde^<_m(t,t')
\\
&=2\zeta'\heav{t'-t}f(t,t'),
\end{split}
\end{equation}
and anti-time-ordered ($(z,z')\in \sC^-\times\sC^-$),
\begin{equation}\label{eq:KrmTB_micro}
K^\rmTb(t,t')=2\zeta'\heav{t-t'}f(t,t'),
\end{equation}
components of the memory kernel $K(z,z')$.

In the Markovian limit, $f(t,t')\to\dirac{t-t'}$ and $\heav{t-t'}f(t-t') \to(1/2)\,\delta(t-t'-0^+)$ and the memory kernel derived from the microscopic theory coincides with the one in Eqs.~(\ref{eq:SM_K_kernel_T})--(\ref{eq:SM_K_kernel_>}).

\subsection{Non-Markovian thermal bath}
To conclude this Appendix, we present an example in which the system of interest is coupled to an external SYK thermal bath. Similar setups have been considered in Refs.~\cite{chen2017JHEP,zhang2019PRB,almheiri2019ARXIV,zhou2020PRB,maldacena2021JHEP,cheipesh2021PRB,haldar2020PRR,su2021JHEP,chen2021JHEP,zanoci2022PRR}. In general, the dynamics will be non-Markovian, although the Markovian approximation may be recovered in some limits. 

We again consider the system to be composed of $N$ Majorana operators $\chi_i$, but now explicitly couple it to a bath of $N_\mathrm{B}$ Majorana fermions $\psi_a$, $a=1,\dots,N_\mathrm{B}$, with $N\ll N_\mathrm{B}$. Within the same setup of the previous sections, we take $\hat{X}_m=L_m$, with $L_m$ given by Eq.~(\ref{eq:H_L_SYK_def}), and a quadratic coupling to the bath,
\begin{equation}
    \hat{Y}_m=\i\sum_{a<b}^{N_\mathrm{B}} v_{m,ab}\psi_a\psi_b,
\end{equation}
where $v_{m,ab}$ is a Gaussian random variable with zero mean and variance
\begin{equation}
    \av{\abs{v_{m,ab}}^2}=\frac{v^2}{N_\mathrm{B}^2}.
\end{equation}
The system-reservoir coupling, Eq.~(\ref{eq:coupling_SR}), then reads as 
\begin{equation}\label{eq:HSR_SYK}
    H_\mathrm{SR}=\sum_{i<j}^N\sum_{a<b}^{N_\mathrm{B}} \sum_{m=1}^M \ell_{m,ij}v_{m,ab}^*\chi_i\chi_j\psi_a\psi_b
    +\mathrm{h.c.}
\end{equation}
It is straightforward to include in $H_\mathrm{SR}$ more system Majoranas $\chi_i$ (which would lead to higher-than-quadratic jump operators, see Appendix~\ref{app:generalizations}) or bath Majoranas $\psi_a$, but for definiteness, we will consider Eq.~(\ref{eq:HSR_SYK}) as it stands.

Under these conditions, the contour correlation function of the environment, Eq.~(\ref{eq:Omega_def}) is given by
\begin{equation}
\begin{split}
    \i\Omega_{\mu\mu'}\left(z,z'\right) &=\av{T_{z}\hat{Y}_{\mu}(z)\hat{Y}_{\mu'}^{\dagger}(z')}
    \\
    &=\sum_{\substack{a<b\\c<d}}^{N_\mathrm{B}}\av{v_{\mu,ab}v^*_{\mu,cd}}\delta_{\mu\mu'}
    \av{T_z\psi_a(z)\psi_{b}(z)\psi_{d}(z')\psi_{c}(z')}
    \\
    &=\delta_{\mu\mu'}\frac{v^2}{N_\mathrm{B}^2}\sum_{a<b}^{N_\mathrm{B}}
    \av{T_z\psi_a(z)\psi_{a}(z')\psi_{b}(z)\psi_{b}(z')}.
\end{split}
\end{equation}
If the bath Hamiltonian is also an SYK model, at its saddle-point, the correlation function $\i\Omega$ coincides with the square of the collective field $G_\mathrm{B}(z,z')$, the bath Green's function, defined as
\begin{equation}
    G_{\mathrm{B}}(z,z')=-\frac{\i}{N_\mathrm{B}}\sum_{a=1}^{N_\mathrm{B}}f_a(z)f_a(z'),
\end{equation}
where $f_a(z)$ is a real Grassmann variable on the Keldysh contour representing the action of the bath Majorana operator $\psi_a$. More precisely, we have
\begin{equation}
    \i\Omegatilde_\mu(z,z')=
    \frac{v^2}{2} [\, \i G_\mathrm{B}(z,z')]^2,
\end{equation}
where $\Omegatilde$ was defined in Eq.~(\ref{eq:Omega_restricted}). Then, from Eq.~(\ref{eq:micro_def_K}), it follows that the kernel reads as (we have $\zeta'=1$ for quadratic operators)
\begin{equation}\label{eq:non-Markovian_kernel}
    K(z,z')=v^2[\,\i G_\mathrm{B}(z,z')]^2.
\end{equation}
As mentioned above, this kernel is manifestly non-Markovian. If $N_\mathrm{B}\gg N$, the dynamics of the bath decouple from the system and can be solved for independently. We can then use the bath Green's function as an input for the Liouvillian theory with jump operators $\hat{X}_m=L_m$. For example, if the bath is an SYK operator in thermal equilibrium at inverse temperature $\beta$, its lesser Green's function reads as [using Eqs.~(\ref{eq:SM_G<_rho}), (\ref{eq:SM_GR_rho}), (\ref{eq:FDT}), and (\ref{eq:SM_numerics_equi_GR}) below]
\begin{equation}
    \i G^<_\mathrm{B}(\omega)
    =-\frac{\beta^{1/2}}{(4\pi)^{1/4}}\frac{2}{e^{\beta \omega}+1}\mathrm{Re}\,\frac{
    \bm{\Gamma}\!\(\frac{1}{4}-\i \frac{\beta\omega}{2\pi}\)}{
    \bm{\Gamma}\!\(\frac{3}{4}-\i \frac{\beta\omega}{2\pi}\)},
\end{equation}
where $\bm{\Gamma}$ is the Gamma function and we have implicitly assumed the bath to have infinite bandwidth. Fourier transforming back to the time-domain, we find
\begin{equation}
    [\i G_\mathrm{B}^<(t)]^2=
    \frac{\beta}{\sqrt{4\pi}}
    \int \frac{\d \omega}{2\pi}\frac{\d \nu}{2\pi}
    \frac{4e^{-\i \omega t}}{(e^{\beta (\omega-\nu)}+1)(e^{\beta \nu}+1)}
    \,\mathrm{Re}\,\frac{
    \bm{\Gamma}\!\(\frac{1}{4}-\i\frac{\beta(\omega-\nu)}{2\pi}\)}{
    \bm{\Gamma}\!\(\frac{3}{4}-\i\frac{\beta(\omega-\nu)}{2\pi}\)}
    \mathrm{Re}\,\frac{
    \bm{\Gamma}\!\(\frac{1}{4}-\i\frac{\beta\nu}{2\pi}\)}{
    \bm{\Gamma}\!\(\frac{3}{4}-\i\frac{\beta\nu}{2\pi}\)}.
\end{equation}
Finally, for large temperatures, $\beta\ll1$, and with the judicious choice of bath variance
\begin{equation}
    v^2=2\,\frac{(4\pi)^{1/4}}{\beta^{1/2}}
    \frac{
    \bm{\Gamma}\!\(\frac{1}{4}\)}{
    \bm{\Gamma}\!\(\frac{3}{4}\)},
\end{equation}
the memory kernel follows from Eq.~(\ref{eq:non-Markovian_kernel}) [using Eqs.~(\ref{eq:SM_G^>})--(\ref{eq:SM_G^Tb}) below],
\begin{equation}
    K^<(t-t')=K^>(t-t')=K^\rmT(t-t')=K^\rmTb(t-t')=\delta(t-t'),
\end{equation}
which, for Hermitian jump operators, is equivalent to the Markovian memory kernel of Eqs.~(\ref{eq:SM_K_kernel_T})--(\ref{eq:SM_K_kernel_>}).

\section{Derivation of the effective action for collective fields}
\label{app:details_effective_action}

In this Appendix, we fill in the details of the derivation of the mean-field action for the collective fields $(G,\Sigma)$, Eq.~(\ref{eq:C_S_eff}), outlined in the Main Text. In particular, we comment on the choice of scaling of the couplings.

We are interested in the averaged partition function,
\begin{align}
\av{Z}&=\int \prod_{i=1}^N\sD a_i\
\av{e^{\i S[a_i]}}_{J,\Gamma},
\\
\label{eq:SM_C_Keldysh_action}
\begin{split}
\i S[a_i]&=
\i\int_\sC \d z\,\frac{1}{2} \sum_{i=1}^N
a_{i}(z)\,\i\pd_za_{i}(z)
-\i \int_\sC \d z\sum_{i<j<k<l}^N
J_{ijkl}a_i(z)a_j(z)a_k(z)a_l(z)
\\
&+\int_\sC \d z \d z'\,K(z,z')\sum_{\substack{i<j\\k<l}}^N
\Gamma_{ijkl} a_i(z)a_j(z)a_k(z')a_l(z').
\end{split}
\end{align}
where the average is performed over both the unitary and dissipative disorder (i.e., over $J_{ijkl}$ and $\Gamma_{ijkl}$, respectively). The disorder average of the unitary contribution to the action is straightforward because the random variables $J_{ijkl}$ are, as usual, chosen Gaussian with mean and variance
\begin{equation}\label{eq:SM_J_moments}
\av{J_{ijkl}}=0
\quad\text{and}\quad
\av{J_{ijkl}^2}=\frac{3!J^2}{N^3},
\end{equation}
respectively.
Then, the averaged unitary contribution to the path integral reads as
\begin{equation}\label{eq:av_Herm_Ham}
\begin{split}
&\av{
	\exp\left\{-\i \int_\sC \d z\sum_{i<j<k<l}^N
	J_{ijkl}a_i(z)a_j(z)a_k(z)a_l(z)\right\}
}_J
\\
&=\exp\left\{
-\frac{1}{2}\int_\sC\d z\d z' \sum_{i<j<k<l}^N
\av{J_{ijkl}^2}
a_i(z)a_j(z)a_k(z)a_l(z)a_i(z')a_j(z')a_k(z')a_l(z')
\right\}
\\
&=
\exp\left\{
-\frac{1}{2}\int_\sC \d z \d z'\, \frac{3!J^2}{N^3}\frac{1}{4!}\(\sum_{i=1}^N a_i(z)a_i(z')\)^4
\right\}
\\
&=\exp\left\{-\frac{NJ^2}{8}\int_\sC \d z \d z' 
\left[G(z,z')\right]^4
\right\},
\end{split}
\end{equation}
where we used the definition of the (mean-field) Green's function:
\begin{equation}\label{eq:SM_G_def}
G(z,z')=-\frac{\i}{N}\sum_{i=1}^N  a_{i}(z) a_{i}(z').
\end{equation}

The disorder average of the dissipative contribution cannot be carried out in full generality since the action is not linear in the random variables $\ell_{m,ij}$. Nonetheless, if the number of decay channels $M$ is large, then, using Eq.~(\ref{eq:Gamma_def}) and the Central Limit Theorem (CLT), the random variables $\Gamma_{ijkl}$ become Gaussian-distributed with nonzero mean. Therefore, we disorder-average over $\Gamma_{ijkl}$ instead of over $\ell_{m,ij}$.
(If working in a perturbative approach, this can be understood as a two-loop computation of the averaged partition function. The one-loop computation, i.e., quadratic order in $\ell_{m,ij}$, takes only the mean of $\Gamma_{ijkl}$ into account, and not its variance, and corresponds to evolution under an average Liouvillian.)
By choosing the mean and variance of the independent Gaussian variables $\ell_{m,ij}$ to be
\begin{equation}
\av{\ell_{m,ij}}=0
\quad\text{and}\quad
\av{\abs{\ell_{m,ij}}^2}=\sigma_\ell^2,
\end{equation}
and assuming $M$ is large (the exact scaling of $M$ and $\sigma_\ell^2$ with $N$ will be determined consistently below), it follows from Eq.~(\ref{eq:Gamma_def}) that the only nonzero mean and variance of $\Gamma_{ijkl}$ are
\begin{equation}\label{eq:SM_Gamma_moments}
\av{\Gamma_{ijij}}=M\sigma_\ell^2
\quad\text{and}\quad
\av{\abs{\Gamma_{ijkl}}^2}_\mathrm{con}
=\av{\abs{\Gamma_{ijkl}}^2}-
\abs{\av{\Gamma_{ijkl}}}^2=
M\sigma_\ell^4,
\end{equation}
respectively.
Under these conditions, the dissipative contribution to the averaged partition function reads as
\begin{equation}\label{eq:av_nonHerm_Ham}
\begin{split}
&\av{\exp\left\{
	\int_\sC \d z \d z'\,K(z,z') \sum_{\substack{i<j\\k<l}}^N
	\Gamma_{ijkl} a_i(z)a_j(z)a_k(z')a_l(z')
	\right\}}_\Gamma\\
&=\exp\left\{
\int_\sC \d z \d z'\,K(z,z')
\sum_{i<j}^N
\av{\Gamma_{ijij}} a_i(z)a_j(z)a_i(z')a_j(z')
\right\}\\
&\times	
\exp\left\{\frac{1}{2}\int_\sC \d z \d z' \d w \d w'\,
K(z,z') K(w,w')
\sum_{\substack{i<j\\k<l}}^N
\av{\Gamma_{ijkl}\Gamma_{klij}}_\mathrm{con} a_i(z)a_j(z)a_k(z')a_l(z')a_k(w)a_l(w)a_i(w')a_j(w')
\right\}\\
&=\exp\left\{
\frac{M\(N\sigma_\ell\)^2}{2} \int_\sC \d z \d z'\, K(z,z') \left[G(z,z')\right]^2\right\}\\
&\times
\exp\left\{\frac{M\(N\sigma_\ell\)^4}{8}
\int_\sC \d z \d z' \d w \d w'\,
K(z,z') K(w,w')
\left[G(z,w')\right]^2\left[G(z',w)\right]^2
\right\}.
\end{split}
\end{equation}

We can now choose the scalings of $M$ and $\sigma_\ell^2$ such that (i) all three contributions to the action (unitary, one-loop dissipative, and two-loop dissipative) have the same (linear) scaling with $N$ and (ii) the CLT is applicable. These conditions are uniquely satisfied if we set
\begin{equation}\label{eq:dissipative_scalings}
M=mN
\quad\text{and}\quad
\sigma_\ell^2=\frac{\gamma^2}{N^2}.
\end{equation}

Next, we make the Green's function dynamical by enforcing the definition~(\ref{eq:SM_G_def}) in the path integral. We introduce the self-energy, $\Sigma(z,z')=-\Sigma(z',z)$ conjugate to $G(z,z')$. Then, using the integral representation of the matrix Dirac delta and the associated resolution of the identity,
\begin{equation}
\begin{split}
1&=\int \sD G \dirac{G(z,z')+\frac{\i}{N}\sum_{i}a_i(z)a_i(z')}
\\
&=\int \sD G \sD \Sigma \exp\left\{
-\frac{N}{2}\int_\sC \d z \d z'\, \Sigma(z,z')\(
G(z,z')+\frac{\i}{N}\sum_{i}a_i(z)a_i(z')
\)\right\},
\end{split}
\end{equation}
the averaged partition function reads as
\begin{equation}
\av{Z}=\int \sD G \sD \Sigma e^{\i S_1[G,\Sigma]}
\(
\int \sD a\, e^{\i S_2[a,\Sigma]}
\)^N,
\end{equation}
with actions
\begin{equation}\label{eq:S_1}
\begin{split}
\i S_1[G,\Sigma]=
\frac{N}{2}&\(
-\frac{J^2}{4}\int_\sC \d z \d z' \left[G(z,z')\right]^4
+\frac{m\gamma^4}{4}
\int_\sC \d z \d z' \d w \d w'\,
K(z,z') K(w,w')
\left[G(z,w')\right]^2\left[G(z',w)\right]^2
\right.\\
&\left.
\ +\,m\gamma^2 \int_\sC \d z\, K(z,z') \left[G(z,z')\right]^2
- \int_\sC \d z \d z'\,\Sigma(z,z')G(z,z')
\)
\end{split}
\end{equation}
and
\begin{equation}\label{eq:S_2}
\i S_2[a,\Sigma]=
-\frac{1}{2} \int_\sC \d z \d z'
a(z)\(\dirac{z-z'}\pd_z+\i \Sigma(z,z')\)a(z').
\end{equation}
The action~(\ref{eq:S_2}) is quadratic in the Grassmann fields, which can, therefore, be integrated out. This integration yields $\(\det[\pd+\i\Sigma]\)^{N/2}\propto \(\det[\i\pd-\Sigma]\)^{N/2}$. We finally arrive at the effective action for $(G,\Sigma)$ on the time-contour $\sC$, as stated in Eq.~(\ref{eq:C_S_eff}):
\begin{equation}\label{eq:SM_C_S_eff}
\begin{split}
\i &S_\eff[G,\Sigma]=
\,\frac{N}{2}\Bigg(
\Tr\log\(\i\pd-\Sigma\)
-\int_\sC \d z \d z'\, \Sigma(z,z') G(z,z')
-\frac{J^2}{4}\int_\sC \d z\d z'\, \left[G(z,z')\right]^4
\\
&+\frac{m\gamma^4}{4}\int_\sC \d z \d z' \d w \d w' \,
K(z,z')K(w,w')
\left[G(z,w')\right]^2 \left[G(z',w)\right]^2
+m\gamma^2\int_\sC \d z \d z' \,
K(z,z')\left[G(z,z')\right]^2
\Bigg).
\end{split}
\end{equation}

\section{Numerical solution of the Schwinger-Dyson equations}
\label{app:numerics}

In this Appendix, we discuss in detail the solution of the Schwinger-Dyson equations. We start by defining the different components of the mean-field Green's functions and the relations between them, focusing, in particular, on the case of Majorana fermions. We then discuss in more detail the Keldysh rotation performed in the Main Text and derive Eqs.~(\ref{eq:SDE_sigma+-}) and (\ref{eq:SDE_rho+-}) (the Schwinger-Dyson equations for $\rho^\pm$ and $\sigma^\pm$). Then, we elaborate on the numerical method used to solve these equations. We also present numerical results for thermal equilibrium. Finally, we discuss the regime of validity of the self-consistent Lorentzian ansatz for the spectral function, Eq.~(\ref{eq:rho-_ansatz_largegamma}).

\subsection{Majorana Green's functions and self-energies}
\label{app:Green's_functions}

The Keldysh-contour Green's function $\mathcal{G}$ is defined as the contour-ordered two-point correlation function
\begin{equation}
    \mathcal{G}_{ij}(z,z')=-\i \av{T_z\, \chi_i(z) \chi_j(z')}.
\end{equation}
At the saddle point of the SYK model (i.e., at the mean-field level) it coincides with the collective field $G$ defined in Eq.~(\ref{eq:G_def}) of the Main Text,
\begin{equation}
    G(z,z')=-\frac{\i}{N}\sum_{i=1}^N a_i(z) a_i(z').
\end{equation}
From now on, we assume that we are at the saddle-point, hence $G\,\delta_{ij}=\mathcal{G}_{ij}$ and we refer to $G$ as the Green's function. 

We obtain the different components of the Green's function in real time by restricting $(z,z')$ to the two branches of the contour, $\sC^+$ and $\sC^-$. We begin by defining the greater and lesser Majorana Green's functions,
\begin{align}\label{eq:SM_G^>}
    &G^>(t_1,t_2)=
    G(t_1^-,t_2^+)=
    -\frac{\i}{N}\sum_{i=1}^N a_i(t_1^-) a_i(t_2^+)
\intertext{and}
\label{eq:SM_G^<}
    &G^<(t_1,t_2)=
    G(t_1^+,t_2^-)=
    -\frac{\i}{N}\sum_{i=1}^N a_i(t_1^+) a_i(t_2^-)
    =-G^>(t_2,t_1)
\end{align}
where, as before, $a_i(t^+)$ is a Majorana fermion propagating forward in time (along the contour $\sC^+$) and $a_i(t^-)$ propagates backward (along $\sC^-$). Unlike for complex fermions, only one of $G^>$ and $G^<$ is independent. Next, we have the time-ordered Green's function,
\begin{equation}\label{eq:SM_G^T}
\begin{split}
    G^\rmT(t_1,t_2)&=
    G(t_1^+,t_2^+)=
    -\frac{\i}{N}\sum_{i=1}^N a_i(t_1^+) a_i(t_2^+)
    \\
    &=\heav{t_1-t_2}G^>(t_1,t_2)+\heav{t_2-t_1}G^<(t_1,t_2)
    \\
    &=\heav{t_1-t_2}G^>(t_1,t_2)-\heav{t_2-t_1}G^>(t_2,t_1),
\end{split}
\end{equation}
and the anti-time-ordered Green's function,
\begin{equation}\label{eq:SM_G^Tb}
\begin{split}
    G^\rmTb(t_1,t_2)&=
    G(t_1^-,t_2^-)=
    -\frac{\i}{N}\sum_{i=1}^N a_i(t_1^-) a_i(t_2^-)
    \\
    &=\heav{t_2-t_1}G^>(t_1,t_2)+\heav{t_1-t_2}G^<(t_1,t_2)
    \\
    &=\heav{t_1-t_2}G^<(t_1,t_2)-\heav{t_2-t_1}G^<(t_2,t_1).
\end{split}
\end{equation}
The time-ordered and anti-time ordered components are also related to the greater and lesser components through the identity
\begin{equation}
    [G^\rmT(t_1,t_2)]^2+[G^\rmTb(t_1,t_2)]^2=[G^>(t_1,t_2)]^2+[G^<(t_1,t_2)]^2,
\end{equation}
of which we made use in deriving the saddle-point equation for the greater self-energy, Eq.~(\ref{eq:Sigma>}).

Another useful set of Green's functions (obtained after a Keldysh rotation) are the retarded Green's function,
\begin{equation}\label{eq:SM_G^R}
\begin{split}
    G^\rmR(t_1,t_2)
    &=\heav{t_1-t_2}\(G^>(t_1,t_2)-G^<(t_1,t_2)\)
    \\
    &=\heav{t_1-t_2}\(G^>(t_1,t_2)+G^>(t_2,t_1)\),
\end{split}
\end{equation}
the advanced Green's function,
\begin{equation}\label{eq:SM_G^A}
\begin{split}
    G^\rmA(t_1,t_2)
    &=\heav{t_2-t_1}\(G^<(t_1,t_2)-G^>(t_1,t_2)\)
    \\
    &=-\heav{t_2-t_1}\(G^>(t_1,t_2)+G^>(t_2,t_1)\)
    \\
    &=-G^\rmR(t_2,t_1),
\end{split}
\end{equation}
and the Keldysh Green's function,
\begin{equation}\label{eq:SM_G^K}
\begin{split}
    G^\rmK(t_1,t_2)
    &=G^>(t_1,t_2)+G^<(t_1,t_2)
    \\
    &=G^>(t_1,t_2)-G^>(t_2,t_1)
    \\
    &=-G^\rmK(t_2,t_1).
\end{split}
\end{equation}
The components of the contour self-energy $\Sigma(z,z')$ satisfy the same relations as in Eqs.~(\ref{eq:SM_G^<})--(\ref{eq:SM_G^K}).

If the Green's functions depend only on the difference $t=t_1-t_2$, as in the long-time limit discussed in the Main Text, then we can move to Fourier space, using the convention
\begin{equation}
\begin{split}
A(t)&=\int \frac{\d\omega}{2\pi}\, \hat{A}(\omega)\,e^{-\i \omega t},
\\
\hat{A}(\omega)&=\int \d t\, A(t)\,e^{\i \omega t},
\end{split}
\end{equation}
for a function $A(t)$ and its Fourier transform $\hat{A}(\omega)$ (we omit the hat on the Fourier transform whenever no confusion arises). For reference, in this convention, the Fourier transform of the step function is 
\begin{equation}
    \hat{\Theta}(\omega)=-\sP \frac{1}{\i \omega}+\pi\dirac{\omega}.
\end{equation}
Next, we introduce the quantities
\begin{align}
\label{eq:SM_rho^+-_def}
\rho^{\pm}(\omega)
&=-\frac{1}{2\pi \i}\(G^>(\omega)\pm G^<(\omega)\)
=-\frac{1}{2\pi \i}\(G^>(\omega)\mp G^>(-\omega)\),
\\
\label{eq:SM_sigma^+-_def}
\sigma^{\pm}(\omega)
&=-\frac{1}{2\pi \i}\(\Sigma^>(\omega)\pm \Sigma^<(\omega)\)
=-\frac{1}{2\pi \i}\(\Sigma^>(\omega)\mp \Sigma^>(-\omega)\),
\end{align}
and their Hilbert transforms
\begin{align}
\label{eq:SM_rhoH_def}
\rho^\rmH(\omega)&=-\frac{1}{\pi}\sP\!\int \d \nu\, \frac{\rho^-(\nu)}{\omega-\nu},
\\
\label{eq:SM_sigmaH_def}
\sigma^\rmH(\omega)&=-\frac{1}{\pi}\sP\!\int \d \nu\, \frac{\sigma^-(\nu)}{\omega-\nu}.
\end{align}
We can now rewrite all the components of the Green's function in terms of $\rho^\pm(\omega)$. (Exactly the same relations hold for the self-energies $\sigma^\pm$.)
Equation~(\ref{eq:SM_rho^+-_def}) can be immediately inverted to yield the lesser and greater components:
\begin{align}
\label{eq:SM_G>_rho}
    G^{>}(\omega)
    &=-\pi\i\(\rho^+(\omega)+\rho^-(\omega)\),
    \\
\label{eq:SM_G<_rho}
    G^{<}(\omega)
    &=-\pi\i\(\rho^+(\omega)-\rho^-(\omega)\).
\end{align}
The remaining components are found by appropriate convolutions of $G^>(\omega)$ and $\hat{\Theta}(\omega)$. The time-ordered, anti-time-ordered, retarded, advanced, and Keldysh Green's functions are given by, respectively:
\begin{align}
    G^\rmT(\omega)&=-\pi\(\rho^\rmH(\omega)+\i \rho^+(\omega)\)
    \\
    G^\rmTb(\omega)&=\pi\(\rho^\rmH(\omega)-\i \rho^+(\omega)\)
    =-G^\rmT(\omega)^*
    \\
    \label{eq:SM_GR_rho}
    G^\rmR(\omega)&=-\pi\(\rho^\rmH(\omega)+\i \rho^-(\omega)\)
    \\
    G^\rmA(\omega)&=-\pi\(\rho^\rmH(\omega)-\i \rho^-(\omega)\)
    =G^\rmR(\omega)^*
    \\
    G^\rmK(\omega)&=-2\pi \i\rho^+(\omega).
\end{align}

\subsection{Derivation of Eqs.~(\ref{eq:SDE_sigma+-}) and (\ref{eq:SDE_rho+-})}

The Schwinger-Dyson equation for the self-energy, Eq.~(\ref{eq:Sigma>}), is given by multiplication in time and hence a convolution in Fourier space. Straightforward algebraic manipulation using Eqs.~(\ref{eq:SM_G>_rho}) and (\ref{eq:SM_G<_rho}) immediately gives
\begin{equation}\label{eq:SM_SDE_sigma+-}
\sigma^\pm(\omega)=
\frac{(1\mp1)m\gamma^2}{2\pi}
+\frac{1}{4}\int \d\mu \d\nu\, \rho^\pm(\omega-\mu-\nu) \left[
\(J^2+m\gamma^4\) \rho^\pm(\mu)\rho^\pm(\nu)
+\(3J^2-m\gamma^4\)\rho^\mp(\mu)\rho^\mp(\nu)
\right].
\end{equation}
Setting $\rho^+(\omega)=\sigma^+(\omega)=0$ in the preceding equation leads to Eq.~(\ref{eq:SDE_sigma+-}).

To obtain Eq.~(\ref{eq:SDE_rho+-}) we start from the Dyson equation on the Keldysh contour, Eq.~(\ref{eq:C_SD_Sigma}), written as
\begin{equation}\label{eq:SM_Dyson}
    \int_\sC \d z'' \(
    G_0^{-1}(z,z'')-\Sigma(z,z'')
    \)G(z'',z')=
    \delta(z-z'),
\end{equation}
where $G_0^{-1}(z,z')=\i\delta(z-z')\pd_z$.
Restricting $(z,z')$ to $\sC^+\times\sC^-$ and $\sC^-\times\sC^+$ (i.e., applying Langreth's rules) the right-hand side of Eq.~(\ref{eq:SM_Dyson}) vanishes and, noting also that $G_0^{-1>}=G_0^{-1<}=0$, we obtain, in Fourier space,
\begin{align}
    \Sigma^>(\omega)G^\rmA(\omega)=\(\omega-\Sigma^\rmR\)G^>(\omega),\\
    \Sigma^<(\omega)G^\rmA(\omega)=\(\omega-\Sigma^\rmR\)G^<(\omega).
\end{align}
Taking the sum and the difference of these equations, we can relate $\rho^\pm$ with $\sigma^\pm$:
\begin{equation}\label{eq:SM_SDE_rho+-}
\begin{split}
    \rho^\pm(\omega)
    &=\sigma^\pm(\omega)\,
    \frac{G^\rmA(\omega)}{\omega-\Sigma^\rmR(\omega)}
    =\sigma^\pm(\omega)\,
    \frac{G^\rmA(\omega)\(\omega-\Sigma^\rmA(\omega)\)}{\abs{\omega-\Sigma^\rmR(\omega)}^2}
    \\
    &=\frac{\sigma^\pm(\omega)}{
    \(\omega+\pi \sigma^\rmH(\omega)\)^2
    +\(\pi \sigma^-(\omega)\)^2},
\end{split}
\end{equation}
where we used the equivalent of Eqs.~(\ref{eq:SM_G^R}) and (\ref{eq:SM_G^A}) for the self-energy and the fact that the Dyson equation is diagonal for the retarded and advanced components, $(\omega-\Sigma^\rmA)G^\rmA=(\omega-\Sigma^\rmR)G^\rmR=1$.

\subsection{Numerical method}

We solved Eqs.~(\ref{eq:SDE_sigma+-}) and (\ref{eq:SDE_rho+-}) iteratively on a linearly-discretized frequency grid $\Lambda=\{-\omega_\mathrm{\max},-\omega_\mathrm{\max}+\Delta\omega,\dots,\omega_\mathrm{\max}\}$, with $\omega_\mathrm{max}=1000$ and $\Delta\omega=0.05$. We used the following procedure:
\begin{enumerate}
    \item Given a spectral function $\rho^-_i(\omega)$, we compute a new self-energy $\sigma_{i+1}^-(\omega)$ from Eq.~(\ref{eq:SDE_sigma+-}) by interpolating $\rho^-(\omega)$ [with $\rho^-(\omega\notin\Lambda)=0$] and numerically evaluating the triple convolution.
    \item We evaluate the Hilbert transform $\sigma_{i+1}^\rmH(\omega)$, Eq.~(\ref{eq:def_rhoH}), using the trapezoid rule.
    \item We compute the new spectral function using Eq.~(\ref{eq:SDE_rho+-}). To ensure there is monotone convergence, we do a partial update,
    \begin{equation}
        \rho^-_{i+1}(\omega)=(1-\eta_\mathrm{mix})\,\rho^-_i(\omega)+
        \eta_\mathrm{mix}\,\frac{\sigma_{i+1}^-(\omega)}{
    \(\omega+\pi \sigma_{i+1}^\rmH(\omega)\)^2
    +\(\pi \sigma_{i+1}^-(\omega)\)^2},
    \end{equation}
    with $\eta_\mathrm{mix}=0.1$.
    \item We repeat steps 1.--3.\ until the solution convergences, in the sense that the total difference between two successive iterations is less than some prescribed accuracy, 
    \begin{equation}
        \sum_{\omega\in\Lambda}\abs{\rho^-_{i+1}(\omega)-\rho^-_{i}(\omega)}<\epsilon=10^{-4}.
    \end{equation}
    At each step, we also checked the normalization of the spectral function, $\int \d\omega\, \rho^-(\omega)=1$ to within the prescribed accuracy.
\end{enumerate}
For a given $J$ and $m$, we started from moderately high value of $\gamma$, for which we expect the Lorentzian ansatz, Eqs.~(\ref{eq:rho-_ansatz_largegamma}) and (\ref{eq:analytical_gap}), to be accurate. We used this ansatz as the initial seed for the algorithm outlined above. We then successively lowered $\gamma$ until $\gamma=0$, at each step using a previously converged solution as the new seed. 

As a check on our method, we confirmed that the system indeed relaxes to the infinite-temperature steady state. To that end, we solved Eqs.~(\ref{eq:SM_SDE_sigma+-}) and (\ref{eq:SM_SDE_rho+-}) with a nonzero initial seed for $\rho^+(\omega)$ and found that the equations converged to $\rho^+(\omega)=0$, while $\rho^-(\omega)$ coincides with the solution of Eqs.~(\ref{eq:SDE_sigma+-}) and (\ref{eq:SDE_rho+-}).

To go back to the frequency domain, we computed Eq.~(\ref{eq:GR_rho-}) using the trapezoid rule. With the frequency grid $\Lambda$ described above, we were able to study the time-domain decay of the retarded Green's function down to $\i G^\rmR(t)\sim10^{-6}$.

\subsection{\texorpdfstring{$\gamma=0$}{gamma=0} and finite-temperature equilibrium dynamics}

In this section we consider the singular limit $\gamma=0$. As there is no dissipation, there is no relaxation to a steady state and the dynamics depend on the initial state. If we take the initial state to be in thermal equilibrium at inverse temperature $\beta$, we have the fluctuation-dissipation relation
\begin{equation}\label{eq:FDT}
\rho^{+}(\omega)=\tanh\(\frac{\beta\omega}{2}\)\rho^{-}(\omega),
\end{equation}
which is just a rewriting of
\begin{equation}
    G^\rmK(\omega)=\tanh\(\frac{\beta\omega}{2}\)\(G^\rmR(\omega)-G^\rmA(\omega)\).
\end{equation}
The Schwinger-Dyson equations, Eqs.~(\ref{eq:SM_SDE_sigma+-}) and (\ref{eq:SM_SDE_rho+-}), then read as:
\begin{align}\label{eq:SM_SDE_equi_sigma}
\sigma^-(\omega)&=
\frac{J^2}{4}\cosh\(\beta\omega/2\)\int \d\mu \d\nu \,
\frac{\rho^-(\omega-\mu-\nu)}{\cosh\(\beta(\omega-\mu-\nu)/2\)}
\frac{\rho^-(\mu)}{\cosh\(\beta\mu/2\)}
\frac{\rho^-(\nu)}{\cosh\(\beta\nu/2\)},
\\
\label{eq:SM_SDE_equi_rho}
\rho^-(\omega)&=
\frac{\sigma^-(\omega)}{
\(\omega+\pi \sigma^\rmH(\omega)\)^2
+\(\pi \sigma^-(\omega)\)^2}.
\end{align}
In the Main Text, we enforced the solution $\rho^+(\omega)=0$ at $\gamma=0$ by considering infinite-temperature equilibrium, $\beta=0$.

\begin{figure}[t]
    \centering
    \includegraphics[width=\textwidth]{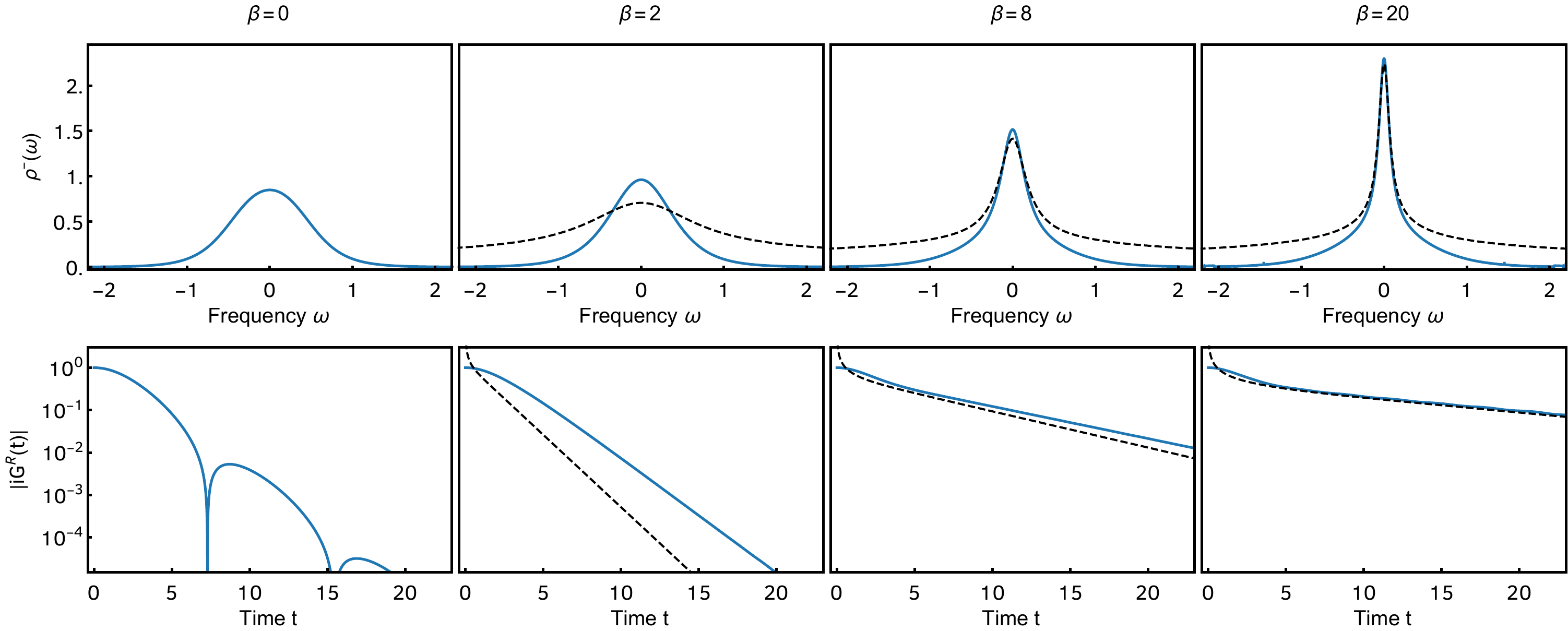}
    \caption{Solution of the equilibrium Schwinger-Dyson equations, Eqs.~(\ref{eq:SM_SDE_equi_sigma}) and (\ref{eq:SM_SDE_equi_rho}), for $J=1$ and different $\beta$. The blue lines show the numerical solution, while the black dashed lines give the conformal ansatz, Eqs.~(\ref{eq:SM_numerics_equi_GRt}) and (\ref{eq:SM_numerics_equi_GR}). Top row: spectral function as a function of frequency. Bottom row: time-decay of the retarded Green's function. We see good agreement of the conformal ansatz with the numerics in the strong-coupling, low-temperature limit, while the $\gamma=0$ limit of the nonequilibrium setup (i.e., at $\beta=0$) falls outside its regime of validity.}
    \label{fig:SM_equilibrium}
\end{figure}

We can benchmark our numerical method by comparing the results for this equilibrium case with the known analytical solution of the SYK model. This solution follows form the (near-)conformal character of the model at low-energies. Besides the UV cutoff, there is also an IR cutoff because of the nonzero temperature, i.e., the conformal solution is accurate for $\beta^{-1} \ll \omega \ll J$.  For the mean-field solution to be exact we further require $N\gg \beta J \gg 1$. Under these conditions, the finite-temperature retarded Green's function is given by~\cite{sachdev2015PRX,maldacena2016PRD}
\begin{equation}\label{eq:SM_numerics_equi_GRt}
    \i G^\rmR(t)=
    \frac{1}{(4\pi)^{1/4}}\sqrt{\frac{2\pi}{\beta}}\frac{1}{\sqrt{\sinh\pi t/\beta}}\heav{t},
\end{equation}
or, in Fourier space, by
\begin{equation}\label{eq:SM_numerics_equi_GR}
    \i G^\rmR(\omega)=
    \frac{1}{\pi^{1/4}}\sqrt{\frac{\beta}{2\pi}}
    \,\bm{B}\!\(\frac{1}{2},\frac{1}{4}-\i \frac{\beta\omega}{2\pi}\)
    =
    \,\pi^{1/4}\sqrt{\frac{\beta}{2\pi}}
    \,\frac{
    \bm{\Gamma}\!\(\frac{1}{4}-\i \frac{\beta\omega}{2\pi}\)}{
    \bm{\Gamma}\!\(\frac{3}{4}-\i \frac{\beta\omega}{2\pi}\)},
\end{equation}
where $\bm{B}$ and $\bm{\Gamma}$ are the Beta and Gamma functions, respectively. Taking the imaginary part of Eq.~(\ref{eq:SM_numerics_equi_GR}) (times $\i/\pi$), we obtain the spectral function $\rho^{-}(\omega)$. In Fig.~\ref{fig:SM_equilibrium} we show the numerical solution of the Schwinger-Dyson equations, Eqs.~(\ref{eq:SM_SDE_equi_sigma}) and (\ref{eq:SM_SDE_equi_rho}), for $J=1$ and different $\beta$. We see that the conformal solution is accurate for low temperatures and small frequencies. Very low temperatures, $\beta\to\infty$, are not accessible to our method, as it is very hard to achieve convergence of the equations. At high temperatures, $\beta \ll J^{-1}$, the conformal solution breaks down. The $\gamma=0$ solution presented in the Main Text falls into this regime.

\subsection{\texorpdfstring{$\gamma^2\gg J$}{gamma2>>J} and the self-consistent Lorentzian approximation}

\begin{figure}[t]
    \centering
    \includegraphics[scale=0.86]{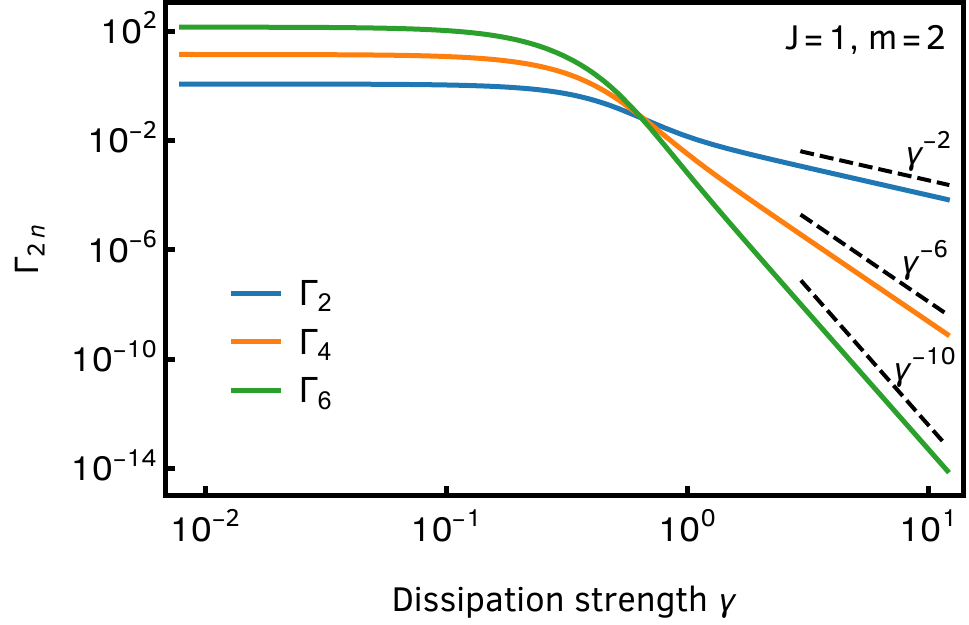}
    \caption{Three lowest-order corrections to the Lorentzian ansatz, $\Gamma_{2n}$ with $n=1,2,3$, as a function of dissipation strength $\gamma$, for $J=1$ and $m=2$. For large dissipation, the corrections have a power-law decay, $\Gamma_{2n}\propto \gamma^{2-4n}$. This supports the claim that the Lorentzian ansatz is exact at $\gamma=\infty$ and very accurate for $\gamma^2\gg J$. At low $\gamma$, the corrections to the ansatz become dominant and the approximation breaks down.}
    \label{fig:SM_Lorentzian_coeffs}
\end{figure}

We now consider the opposite limit $\gamma^2\gg J$ and discuss  the validity of the Lorentzian approximation for the spectral function $\rho^-(\omega)$:
\begin{equation}\label{eq:SM_rho-_ansatz_largegamma}
\rho^-(\omega)=\frac{1}{\pi}\frac{\Gamma}{\omega^2+\Gamma^2}.
\end{equation}
The width $\Gamma$ can be determined self-consistently as follows. Setting $\rho^+(\omega)=0$ and inserting the ansatz of Eq.~(\ref{eq:SM_rho-_ansatz_largegamma}) into Eq.~(\ref{eq:SDE_sigma+-}) we determine $\sigma^-(\omega)$ exactly, from which $\sigma^\rmH(\omega)$ can be computed using Eq.~(\ref{eq:def_rhoH}). Plugging these expressions back into the Dyson equation for $\rho^{-}(\omega)$, Eq.~(\ref{eq:SDE_rho+-}), we find that $\Gamma$ has to satisfy the self-consistency equation
\begin{equation}\label{eq:self_consistency_Gamma}
\Gamma=\frac{
\(\omega^2+\Gamma^2\)
\(m\gamma^2+\frac{J^2+m\gamma^4}{4}\frac{3\Gamma}{\omega^2+9\Gamma^2}\)
}{
\omega^2\(
1+\frac{J^2+m\gamma^4}{4}\frac{1}{\omega^2+9\Gamma^2}
\)^2+\(
m\gamma^2+\frac{J^2+m\gamma^4}{4}\frac{3\Gamma}{\omega^2+9\Gamma^2}
\)^2
}.
\end{equation}
Note that, contrarily to the Main Text, we have not set $\omega=0$. Next, we expand $\Gamma$ in powers of $\omega$, $\Gamma=\Delta+\Gamma_2\omega^2+\Gamma_4\omega^4+\cdots$. At zeroth order in $\omega$, Eq.~(\ref{eq:self_consistency_Gamma})
is solved by
\begin{equation}\label{eq:SM_analytical_gap}
\Delta=\frac{m\gamma^2}{2}\(
1+\sqrt{
\frac{3m+1}{3m}+\frac{J^2}{3m^2\gamma^4}
}
\),
\end{equation}
which is Eq.~(\ref{eq:analytical_gap}). The higher-order coefficients, $\Gamma_2$, $\Gamma_4,\dots$, have to vanish for our ansatz to be consistent, since in deriving the self-consistency equation we implicitly assumed $\Gamma$ to be constant. At each order, by using the solution for lower-order coefficients, we obtain an algebraic equation for the coefficient $\Gamma_{2n}$. For instance, at order $\omega^2$, we use the result at order $\omega^0$ [Eq.~(\ref{eq:SM_analytical_gap})], to find that Eq.~(\ref{eq:self_consistency_Gamma}) yields a linear equation for $\Gamma_2$ solved by
\begin{equation}
\begin{split}
    \Gamma_2=&\,\frac{2}{3m\gamma ^2} \left[\left(\frac{12m^2+9m+1}{9m^2}-\frac{4m+7/3}{3m}\sqrt{\frac{3 m+1}{3m}+\frac{J^2}{3m^2\gamma^4}}\right)\right.
    \\
    +&\left.\,\frac{J^2}{3m^2\gamma^4} \left( \frac{9m+2}{3m}-\frac{7}{3}\sqrt{\frac{3 m+1}{3m}+\frac{J^2}{3m^2\gamma ^4}}\right)
    + \frac{J^4}{9m^4\gamma^8}\right]
    \left[\left(\frac{1}{3m}+\frac{J^2}{3m^2\gamma^4}\right)^2 \sqrt{\frac{3 m+1}{3m}+\frac{J^2}{3m^2\gamma^4}}\right]^{-1},
\end{split}
\end{equation}
which decays as $1/\gamma^2$ as $\gamma\to\infty$. We can proceed analogously for higher-order coefficients. Although it may in principle be possible to show exactly that $\lim_{\gamma\to\infty}\Gamma_{2n}=0$ for all $n$, here we limit ourselves to the first three corrections, $\Gamma_2$, $\Gamma_4$, and $\Gamma_6$, plotted in Fig.~\ref{fig:SM_Lorentzian_coeffs}. This procedure indicates the exactness of the Lorentzian approximation as $\gamma\to\infty$ and also provides a quantitative measure for its accuracy at finite $\gamma$.

\section{SYK Lindbladian with higher-body interactions}
\label{app:generalizations}

In this Appendix, we present the generalization of the four-body SYK Lindbladian of the Main Text for higher $q$-body interactions.

We consider an Hamiltonian consisting of $q$ Majoranas, where $q$ is an even integer, and jump operators with $k$ Majoranas, where $k$ is any positive integer. In the Main Text, we restricted ourselves to $q=2k=4$. The Hamiltonian and jump operators read as
\begin{align}
\label{eq:SMq_H_SYK_def}
H&=\i^{q(q-1)/2}\sum_{i_1<i_2<\cdots<i_q}^{N}J_{i_1i_2\cdots i_q}\chi_{i_1}\chi_{i_2}\cdots \chi_{i_q},
\\
\label{eq:SMq_l_SYK_def}
L_m&=\i^{k(k-1)/2}\sum_{i_1<i_2<\cdots<i_{k}}^{N} \ell_{m,i_1i_2\cdots i_{k}}\chi_{i_1}\chi_{i_2}\cdots \chi_{i_{k}},
\end{align}
respectively, with the variance of the random couplings now given by
\begin{equation}
    \label{eq:SMq_J_l_moments}
    \av{J_{i_1i_2\cdots i_q}^2}=\frac{(q-1)!\,J^2}{N^{q-1}}
    \quad\text{and}\quad
    \av{\abs{\ell_{m,i_1i_2\cdots i_{q/2}}}^2}=\frac{k!\,\gamma^2}{(2k)^{1/2}N^{k}},
\end{equation}
As before, when $N,M\to\infty$ with $m=M/N$ fixed, we have
\begin{equation}\label{eq:SMq_Gamma_moments}
    \av{\Gamma_{i_1i_2\cdots i_{k}i_1i_2\cdots i_{k}}}=\frac{k!\,m\gamma^2}{(2k)^{1/2}N^{k-1}}
    \quad\text{and}\quad
    \av{\abs{\Gamma_{i_1i_2\cdots i_{2k}}}^2}_\mathrm{con}= \frac{(k!)^2 m\gamma^4}{2k N^{2k-1}}.
\end{equation}

Following the same procedure as in Appendix~\ref{app:details_effective_action}, we obtain the $(G,\Sigma)$ effective action,
\begin{equation}\label{eq:SMq_C_S_eff}
\begin{split}
\i &S_\eff[G,\Sigma]=
\,\frac{N}{2}\Bigg(
\Tr\log\(\i\pd-\Sigma\)
-\int_\sC \d z \d z'\, \Sigma(z,z') G(z,z')
-\frac{\i^q J^2}{q}\int_\sC \d z\d z'\, \left[G(z,z')\right]^q
\\
&+\frac{(-1)^k m\gamma^4}{2k}\int_\sC \d z \d z' \d w \d w' \,
K(z,z')K(w,w')
\left[G(z,w')\right]^{k} \left[G(z',w)\right]^{k}
+\frac{\i^{k^2} 2m\gamma^2}{(2k)^{1/2}}\int_\sC \d z \d z' \,
K(z,z')\left[G(z,z')\right]^{k}
\Bigg),
\end{split}
\end{equation}
from which the contour Schwinger-Dyson equations are obtained:
\begin{align}
	\label{eq:SMq_C_SD_Sigma}
	\(\i\pd-\Sigma\)\cdot &\,G=\id_{\sC},
	\\
	\begin{split}\label{eq:SMq_C_SD_G}
	\Sigma(z,z')=&-\i^q J^2\left[G(z,z')\right]^{q-1}
	+(-1)^k \frac{m \gamma^4}{2} \left[G(z,z')\right]^{k-1}
	\int_{\sC} \d w \d w' \,
	\(K(z,w)K(w',z')+(-1)^kK(w,z)K(z',w')
	\)\\
	&\times\left[G(w,w')\right]^{k}
	+\i^{k^2} (k/2)^{1/2}m\gamma^2\(K(z,z')+(-1)^kK(z',z)\)G(z,z')^{k-1}.
	\end{split}
\end{align}
To go to the real-time axis, we use Eqs.~(\ref{eq:SM_K_kernel_T})--(\ref{eq:SM_K_kernel_>}) with $\zeta'=-(-1)^{k(k+1)/2}$. The greater self-energy reads as
\begin{equation}\label{eq:SMq_Sigma>}
\begin{split}
\Sigma^>(t_1,t_2)=
&-\i^q J^2\left[G^>(t_1,t_2)\right]^{q}
\\
&+m\gamma^4\left\{
\frac{1-(-1)^k}{2}
\left[G^>(t_1,t_2)\right]^{2k-1}
-
\left[G^>(t_1,t_2)\right]^{k-1}\left[G^<(t_1,t_2)\right]^{k}
\right\}
\\
&-\i(2k)^{1/2}m\gamma^2\dirac{t_1-t_2}\left[G^>(t_1,t_2)\right]^{k-1}
\end{split}
\end{equation}
and, in terms of the spectral function $\rho^-(\omega)$, as
\begin{equation}
    \sigma^-(\omega)=\frac{m\gamma^2}{\pi}\frac{(2k)^{1/2}}{2^{k-1}}
    +\frac{J^2}{2^{q-2}} (\rho^-)^{*(q-1)}(\omega)
    +\frac{3-(-1)^k}{2}\frac{m\gamma^4}{2^{2k-2}} (\rho^-)^{*(2k-1)}(\omega),
\end{equation}
where $(\rho^-)^{*n}(\omega)$ denotes the $n$-fold convolution of the spectral function with itself,
\begin{equation}
    (\rho^-)^{*n}(\omega)=
    \int \prod_{j=1}^{n-1}\d \nu_j\, 
    \rho^-(\omega-\sum_{j=1}^{n-1} \nu_j)
    \prod_{j=1}^{n-1} \rho^-(\nu_j).
\end{equation}
The spectral gap computed from the Lorentzian ansatz is in turn given by
\begin{equation}
\Delta=m \gamma^2 \frac{(2k)^{1/2}}{2^{k}}\(
1+\sqrt{1+
\frac{(2k-1)km+3-(-1)^k}{(2k-1)km}+\frac{2^{2k-q+1}}{k}\frac{J^2}{(q-1)m^2\gamma^4}
}\).
\end{equation}

\section{Quantum chaos and universality from exact diagonalization results}
\label{app:ed}

In this Appendix, we present some exact diagonalization results for finite $N$ and $M$. We study the spectral statistics, finding excellent agreement with random matrix theory and, therefore, quantum chaotic dynamics, and also compare the full spectrum to that of dense random Lindbladians. While these results are outside the scaling limit and, therefore, are not directly comparable with the rest of the Letter and Supplemental Material, they show there is valuable insight into general chaotic dissipative systems to be gained from the analytical calculations in the SYK Lindbladian model.

\subsection{Local level statistics}

Given that the SYK Lindbladian describes a generic strongly coupled system, it is natural to expect that it is quantum chaotic, i.e., that it displays random matrix statistics. An analytical proof of this statement is a formidable task, that we do not expect to be possible to show directly using the formalism developed in this Letter and Supplemental Material. It can, however, be confirmed numerically by computing complex-spacing-ratio distributions~\cite{sa2019csr}, for which we find excellent agreement with Ginibre random matrix statistics.

\begin{figure}[t]
    \centering
    \includegraphics[width=0.8\textwidth]{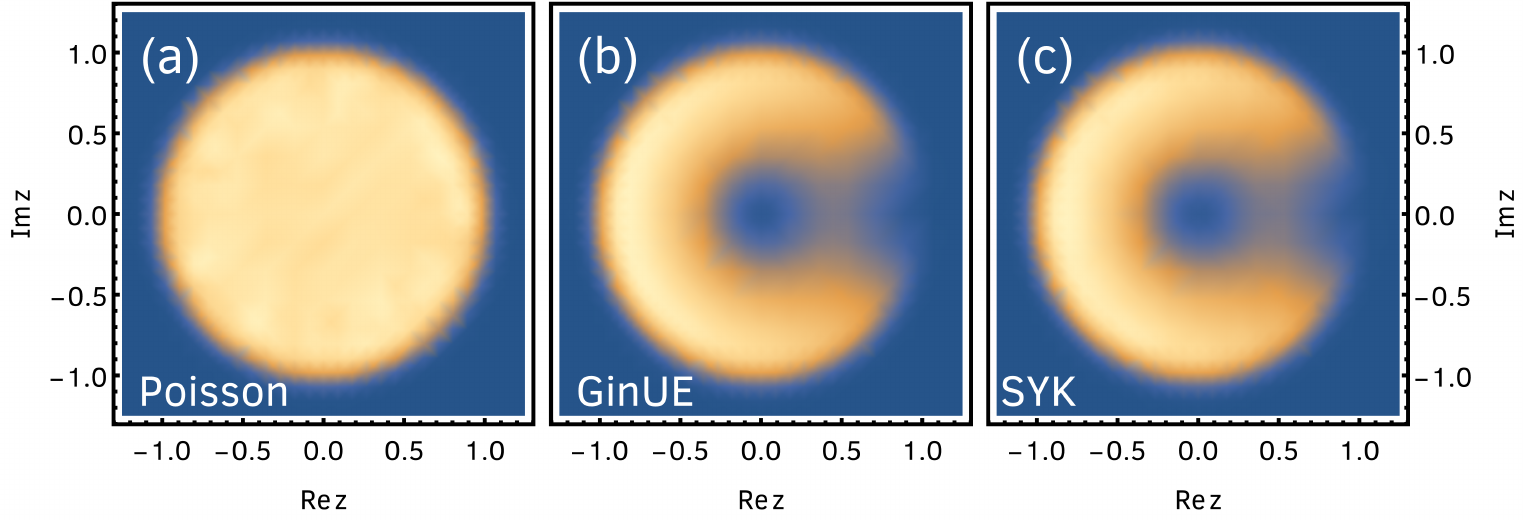}
    \caption{Distribution of the complex spacing ratios, defined in Eq.~(\ref{eq:csr}), for (a) uncorrelated 2d random variables, (b) eigenvalues of random matrices from the Ginibre Unitary Ensemble, and (c) eigenvalues of the SYK Lindbladian for $N=M=16$, $J=1$ and $\gamma=0.5$ (in the sector of even parities). We see that the SYK Lindbladian displays clear random matrix statistics, signaling its quantum chaotic nature. To obtain these plots, we ensemble-averaged over $10^5$ random variables in (a), $256$ random matrices of dimension $32768\times32768$ in (b), and $32$ SYK Lindbladians (each sector has dimension $16384$) in (c).}
    \label{fig:SM_CSR}
\end{figure}

To make the above claim precise, we start by writing down a matrix representation (vectorization) of the Lindbladian of Eq.~(\ref{eq:Lindblad_eq}),
\begin{equation}\label{eq:matrix_rep_L}
    \mathcal{L}=-\i \(H\otimes \id -\id \otimes H^*\)
    +\sum_{m=1}^M \(2 L_m\otimes L_m^*
    -L_m^\dagger L_m\otimes \id
    -\id \otimes L_m^\top L_m^*
    \),
\end{equation}
with $H$ and $L_m$ still given by Eq.~(\ref{eq:H_L_SYK_def}), which can be exactly diagonalized. Since all of $H$ and $L_m$ commute with the fermion parity (chirality) operator, $\gamma_\mathrm{ch}=\i^{N^2/4}\prod_{i=1}^N\chi_i$, the Lindbladian has a Liouvillian strong symmetry~\cite{buca2012} and the matrix representation~(\ref{eq:matrix_rep_L}) block-diagonalizes into four independent sectors of fixed parity of each tensor factor.
Let $\{\Lambda_j\}$, $j=1,\dots,2^{N-2}$, denote the complex eigenvalues of one such block. We then define the complex spacing ratio (CSR)~\cite{sa2019csr} as
\begin{equation}\label{eq:csr}
    z_j=\frac{\Lambda^{\mathrm{NN}}_j-\Lambda_j}{\Lambda^{\mathrm{NNN}}_j-\Lambda_j},
\end{equation}
where $\Lambda^{\mathrm{NN}}$ is $\Lambda_j$'s nearest neighbor and $\Lambda^{\mathrm{NNN}}$ its next-to-nearest neighbor. In the limit $N\to\infty$, the probability distribution of $z_j$ is flat on the unit disk for 2d uncorrelated random variables (Poisson statistics), while for non-Hermitian random matrices it has a characteristic donut-like shape, see Fig.~\ref{fig:SM_CSR}(a)--(b). We applied this prescription to the eigenvalues of a Lindbladian block of positive parities obtained from exact diagonalization of Eq.~(\ref{eq:matrix_rep_L}) for $N=M=16$, $J=1$, and $\gamma=0.5$. By inspection of Fig.~\ref{fig:SM_CSR}(c), we find excellent agreement with random matrix statistics, as expected from the quantum chaos conjecture. A more in-depth comparison (for instance, of the marginal radial and angular distributions) could be performed to check for any quantitative deviations from random matrix statistics, but we do not pursue this matter further here.

\subsection{Global spectral features}

\begin{figure}[t]
    \centering
    \includegraphics[width=0.82\textwidth]{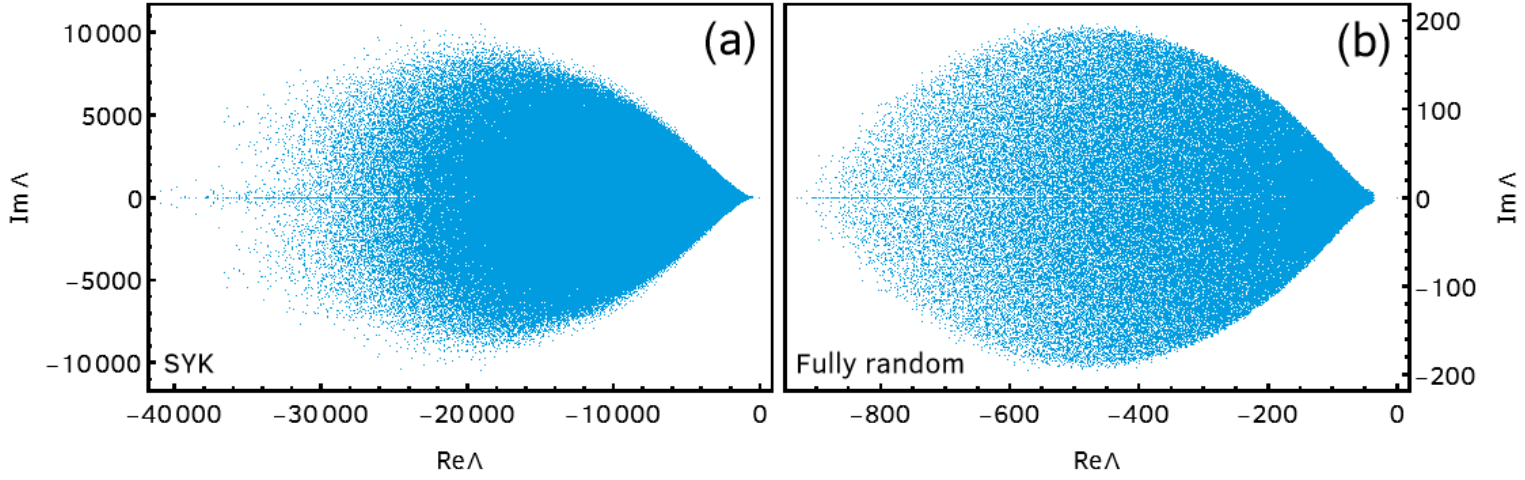}
    \caption{Spectra of random Lindbladians, obtained from exact diagonalization. (a) SYK Hamiltonian and jump operators, with $N=16$, $M=2$, $J=1$, and $\gamma=50$ (in the sector of even parities). (b) Fully random Hamiltonian (sampled from the Gaussian Unitary Ensemble) and two jump operators (sampled from the Ginibre Unitary Ensemble) of dimension $6400\times6400$, with dissipation strength $g_\mathrm{eff}=5$ (defined in Ref.~\cite{sa2019}). Despite the sparseness of the SYK Lindbladian, the spectral support is very similar to the universal lemon-like shape found for dense random Lindbladians~\cite{denisov2018,can2019,sa2019}.}
    \label{fig:SM_spectrum}
\end{figure}

More surprising, perhaps, is the fact that other properties of the SYK Lindbladian, besides local level correlations, are similar to those of other generic models. The full spectrum, which describes not only the asymptotic approach to the steady-state but also the short-lived transient dynamics, is in general not expected to be universal. Nevertheless, the spectrum displayed in Fig.~\ref{fig:SM_spectrum}(a), again obtained from exact diagonalization for $N=16$, $M=2$, $J=1$, and $\gamma=50$, exhibits a spectral support similar to the universal lemon-like shape found for dense random Lindbladians~\cite{denisov2018,sa2019}, see Fig.~\ref{fig:SM_spectrum}(b).

Furthermore, the spectral gap computed in this Letter is qualitatively similar to the gap computed for fully random Lindbladians~\cite{can2019,sa2019}. Indeed, in those references, it was found that, for large dissipation, the gap scales quadratically with the dissipation strength and linearly with the number of jump operators. As dissipation decreases, the gap starts to grow slower than quadratically, but here we can no longer compare results because the weak dissipation and thermodynamic limits do not commute.

\end{document}